\documentclass[sigconf]{acmart}
\AtBeginDocument{%
  }

\usepackage{acmart-taps}
\usepackage{xcolor}
\usepackage{graphicx}
\usepackage{multirow}
\usepackage{booktabs}
\usepackage{enumitem}
\usepackage{fontawesome}
\usepackage{appendix}
\usepackage{lscape}
\usepackage{colortbl}
\usepackage{framed}

\definecolor{shadecolor}{rgb}{0.851, 0.851, 0.851}

\newcommand{\footnoteref}[1]{#1}

\newcommand{\crule}[3][black]{\textcolor{#1}{\rule{#2}{#3}}}

\makeatletter
\newcommand{\thickhline}{%
    \noalign {\ifnum 0=`}\fi \hrule height 1pt
    \futurelet \reserved@a \@xhline
}
\newcolumntype{"}{@{\hskip\tabcolsep\vrule width 1pt\hskip\tabcolsep}}
\makeatother

\aptLtoXcmd{
\def\squaredark{\&\#x25A0;}
\def\squarelight{\&\#x25A1;}
\def\squaremedium{\&\#x25EA;}}
{}

\usepackage{chi25-78}

\setcopyright{acmlicensed}
\copyrightyear{2025}
\acmYear{2025}
\setcopyright{acmlicensed}
\acmConference[CHI '25]{CHI Conference on Human Factors in Computing Systems}{April 26-May 1, 2025}{Yokohama, Japan}
\acmBooktitle{CHI Conference on Human Factors in Computing Systems (CHI '25), April 26-May 1, 2025, Yokohama, Japan}
\acmDOI{10.1145/3706598.3713161}
\acmISBN{979-8-4007-1394-1/25/04}

\begin{document}

\title[Making the Write Connections: Linking Writing Support Tools with Writer Needs]{Making the Write Connections: Linking Writing Support Tools with Writer's Needs}

\author{Zixin Zhao}
\email{zzhao1@cs.toronto.edu}
\orcid{0000-0002-8636-1987}
\affiliation{%
  \institution{Department of Computer Science University of Toronto}
  \city{Toronto}
  \state{Ontario}
  \country{Canada}
}

\author{Damien Masson}
\email{damien.masson@umontreal.ca}
\orcid{0000-0002-9482-8639}
\affiliation{%
  \institution{Department of Computer Science University of Toronto}
  \city{Toronto}
  \state{Ontario}
  \country{Canada}
}
\additionalaffiliation{\institution{Université de Montréal}
  \city{Montreal}
  \country{Canada}
}

\author{Young-Ho Kim}
\email{yghokim@younghokim.net}
\orcid{0000-0002-2681-2774}
\affiliation{%
  \institution{NAVER AI Lab}
  \city{Seongnam}
  \state{Gyeonggi}
  \country{Republic of Korea}
}

\author{Gerald Penn}
\email{gpenn@cs.toronto.edu}
\orcid{0000-0003-3553-8305}
\affiliation{%
  \institution{Department of Computer Science University of Toronto}
  \city{Toronto}
  \state{Ontario}
  \country{Canada}
}

\author{Fanny Chevalier}
\email{fanny@cs.toronto.edu}
\orcid{0000-0002-5585-7971}
\affiliation{%
  \institution{Department of Computer Science and Statistical Sciences University of Toronto}
  \city{Toronto}
  \state{Ontario}
  \country{Canada}
}
\renewcommand{\shortauthors}{Zhao et al.}

\begin{abstract}
This work sheds light on whether and how creative writers' needs are met by existing research and commercial writing support tools (WST). We conducted a need finding study to gain insight into the writers' process during creative writing through a qualitative analysis of the response from an online questionnaire and Reddit discussions on \textit{r/Writing}. Using a systematic analysis of 115 tools and 67 research papers, we map out the landscape of how digital tools facilitate the writing process. Our triangulation of data reveals that research predominantly focuses on the writing activity and overlooks pre-writing activities and the importance of visualization. We distill 10 key takeaways to inform future research on WST and point to opportunities surrounding underexplored areas. Our work offers a holistic and up-to-date account of how tools have transformed the writing process, guiding the design of future tools that address writers' evolving and unmet needs. 
\end{abstract}



\received{12 September 2024}
\received[revised]{13 February 2025}
\received[accepted]{5 June 2025}

\setlength{\abovecaptionskip}{3pt}
\setlength{\belowcaptionskip}{-3pt}

\maketitle
\section{Introduction}


Creative writing is a form of writing that goes beyond technical and professional writing to focus on artistic expression, for example, writing novels and poetry. Many different writing support tools (WST), i.e., digital tools designed to assist writers, have been proposed within both the research and commercial realm to facilitate creative writing. HCI research has explored novel approaches focusing on one specific aspect of the creative writing process at a time, such as exploring character personalities using chatbots~\cite{qin_charactermeet_2024}, authoring interactive narratives~\cite{joyce_storyspace_1991,bernstein_storyspace_2002}, writing on phones~\cite{bonsignore_2013}, and, more recently, co-writing with large language models~\cite{mirowski_co-writing_2023}. Meanwhile, commercial tools typically strive to address the writing workflow holistically, from planning to revising all in one system, as is the case of Scrivener\footnote{https://www.literatureandlatte.com/scrivener/} and SudoWrite~\cite{fang_sudowrite_2024}\footnote{https://www.sudowrite.com/}.

Despite the emergent research and commercial tools seeking to improve current and enable innovative writing workflows, there appears to be a misalignment between the resources being developed and the tools professional writers leverage in their work~\cite{frich_creativity_2018,frich_mapping_2019,frich_twenty_2018}. For example, HCI research prioritizes novelty in research prototypes. Still, it leaves open questions regarding the usefulness and relevance of novel approaches when integrated into actual creative writing practices. Furthermore, research in WST typically identifies a singular aspect of the creative writing process to address, making it difficult to compare the impact of the findings across studies. This may impede the advancement of writing tools as a whole and run the risk of reinventing the wheel.

Previous research~\cite{lee_design_2024,gero_design_2022,wan2024coco} has explored the landscape of WST through literature reviews. These works offer new design spaces~\cite{lee_design_2024,gero_design_2022} and taxonomies~\cite{wan2024coco} that can help inform the design of novel WST. However, these meta-analyses neglect the progress in commercial applications and the evolving needs of writers.
This puts new research at risk of overlooking existing problems and proposing solutions disconnected from writers' realities.

To reconnect research with commercial tools and writers' needs, our work provides a comprehensive view of the WST landscape by triangulating three data sources: academic literature, questionnaire responses, and online discussions on Reddit. We first conducted a literature review on WST specifically for creative writing to examine which needs were addressed by research. Then, we collected responses from an online questionnaire and analyzed discussions from the past five years on a subreddit, \textit{r/Writing}\footnote{https://www.reddit.com/r/writing/}, to gain more insight into the processes and tools used by current writers. We make our annotated dataset of commercial tools and literature available online for future researchers\footnote{https://thewriteconnection.github.io/}. 
Through a deep dive into these sources, our work surfaces ten key insights and unveils new avenues for future research in creative writing. We discover unmet needs related to organization and integrating different mediums into writing practice. One of our main insights is to urge researchers to move the spotlight away from writing activities related to generating text and focus more on other processes like pre-writing, planning, and revising. We also urge more work to explore and diversify visualizations to help creative writing. In the following sections, we describe our methodology, explore our key findings and insights, and discuss potential opportunities for future research.


\section{Recent Surveys on Writing Support Tools}
\label{sec:background}


Research on digital WST had existed since the 1960s before display monitors were paired with keyboards~\cite{kruse_word_2023}. Ever since, surveys have examined the changes digital writing has brought to the writing process~\cite{sharples_designs_1996,hartley_writing_2001,goldberg_effect_2003}. As the field continues to grow, we see the breadth of the surveys narrowing to focus on specific writing environments or contexts.

For example, Strobl et al.~\cite{strobl_digital_2019} surveyed tools within the academic writing context and found a majority of the work focused on support for argumentative essays in English. In contrast, automated support for revision was limited to word or sentence level (e.g. grammar and spelling checks). Others focused on defining a design space for intelligent WST~\cite{gero_design_2022}. Gero et al.~\cite{gero_design_2022} analyzed 30 ACM Digital Library papers from 2017 to 2022 using the cognitive process model of writing~\cite{flower_cognitive_1981} and a taxonomy developed for creativity support tools (CST) at large (whose coverage is dominated by visual, interface, and game design)~\cite{frich_mapping_2019}. Although their work revealed a gap in research related to planning tools with specific goals, their scope was limited because of their focus on work within HCI. Lee et al.~\cite{lee_design_2024} built upon this and contributed an extensively defined design space for intelligent WST. Their goal was to provide a guideline for components to consider when developing such a tool. Similarly, the taxonomy by Wan et al.~\cite{wan2024coco} and typology of Hoque et al.~\cite{hoque_hallmark_2024} both categorize the contributions made by intelligent WST for co-writing tasks.

Weber et al.~\cite{weber_supporting_2023} also used the cognitive process model of writing to define a taxonomy. Unlike prior works, their survey was conducted over several search engines (e.g. IEEE Xplore, ProQuest, ACM, ArXiv, etc.), resulting in 86 relevant papers for analysis. Their taxonomy targeted intelligent WST and gave an overview of the main areas in which WST was developed, including education, scientific writing, administration, and creativity purposes.

Since prior surveys focused on tools within the context of academic writing~\cite{strobl_digital_2019} or provided a broad overview of intelligent systems ~\cite{gero_design_2022,weber_supporting_2023,wan2024coco,lee_design_2024}, we explore previously unexplored contexts to expand the mapped space of WSTs. Moreover, we include commercial tools writers use, enabling a more comprehensive view of current writing practices. Our work complements previous work by looking into existing tools and systems built for creative writing. We strive to offer a holistic and realistic view of the field through an extensive survey of existing works within this research space, cross-mapping various data sources to learn about the writer's needs of today. We use existing design spaces~\cite{lee_design_2024} and taxonomies~\cite{frich_mapping_2019} introduced within HCI to guide our process. 

\begin{table*}[t]
\caption{Inclusion criteria for literature review and examples of paper topics which were excluded}
\small
\begin{tabular}{p{0.17\linewidth}   p{0.45\linewidth} p{0.3\linewidth}}
\hline
\textbf{Criteria}   & \textbf{Inclusion}                & \textbf{Exclusion examples}                  \\ 
\hline
(I) System Presence &
  A tool or feature(s) within an existing tool(s) or strategy for supporting creative writing must be the focus &
  Story planning for presentations, storyboard creation, active text \\[0.1cm]
(II) Target Domain &
  Tool must be for creative writing (e.g. fiction, poetry, theatre script) &
  Tools for language learning, essays, academic papers, legal documents \\[0.1cm]
(III) Research Area & Computer science, writing support & Interactive storytelling (game), air-writing 
    \\[0.1cm]
(IV) Citation Count &
    Papers published before 2019 must have at least 5 citations & 
    \\ \hline
\end{tabular}

\label{tab:inclusion-criteria}
\end{table*}

\begin{figure}[t]
    \centering
    \includegraphics[width=0.7\linewidth]{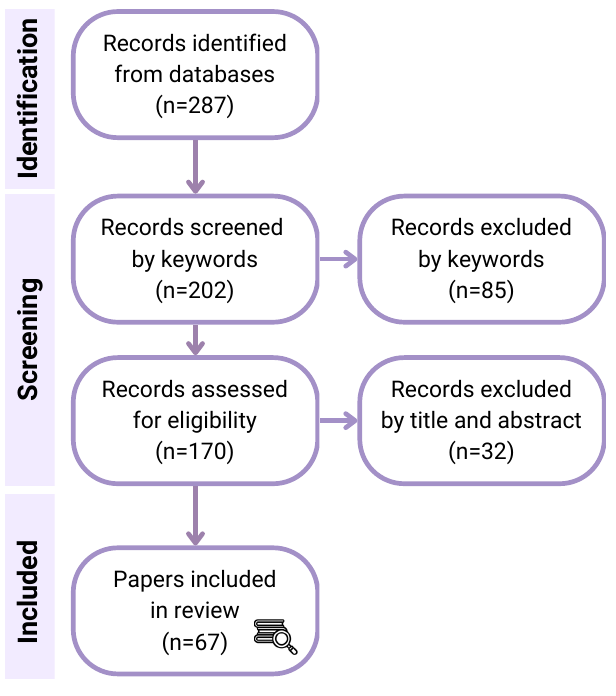}
    \caption{Flow graph of the initial literature search and screening process guided by the PRISMA checklist~\cite{prisma}}
    \label{fig:lit-flow}
    \Description[literature review flow]{litearture review flow diagram for our initial initial literature search}
\end{figure}


\section{Data Sources}
\label{sec:source}


\subsection[]{Literature Sources and Identification \includegraphics[height=0.7\baselineskip]{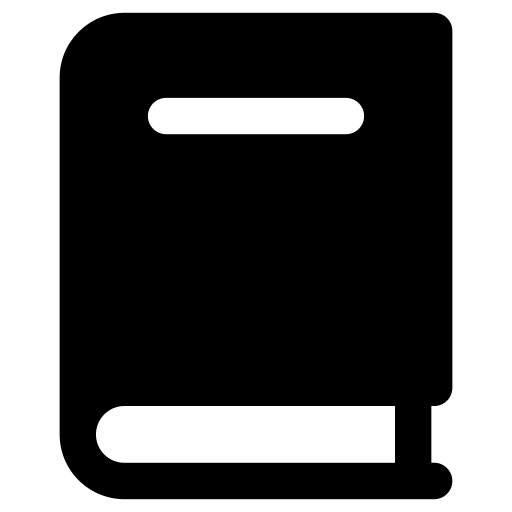}}

\begin{figure*}[t]
    \centering
    \includegraphics[width=\linewidth]{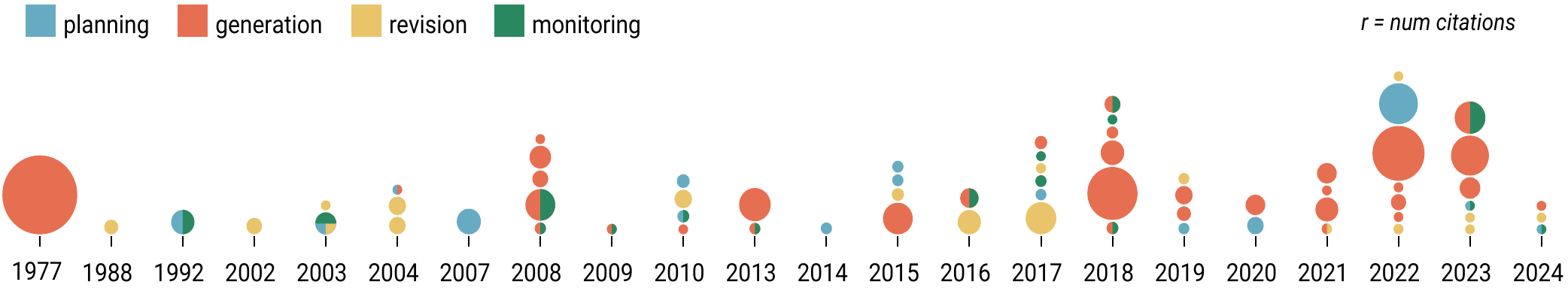}
    \caption{Overview of papers~(n=67) reviewed for our literature survey. Each paper is colour coded with the aspect of writing it targets and spatially positioned by the publication date along the x-axis. The radius is correlated with the number of citations.}
    \Description[Citation Beeswarm]{Stacked circle graph showing on the x-axis years from 1977 to 2024 and in the y-axis the stacked circles representing literature reviewed and number of citation each one has}
    \label{fig:lit-tools}
\end{figure*}

\subsubsection{Initial Literature Search}
\label{sec:prisma}
We conducted a systematic review to investigate the features and objectives of tools proposed in previous works to aid the writer's workflow. We used PRISMA 2020 (Preferred Reporting Items for Systematic Reviews and Meta-Analyses)~\cite{prisma} to guide our process. A team of three coauthors identified five recent representative works and did a broad search over each work's reference to develop an eligibility criteria (\autoref{tab:inclusion-criteria}) and search terms. We searched using the preliminary search terms over ACM Digital Library and met as a group to refine the terms. After four rounds of iterative searches and refinement, we settled on the following search terms over the title and abstracts.

\begin{quote}
(``writer” OR ``author” OR ``writing” OR ``authoring" OR ``post-writing" OR ``editing" OR ``planning" OR ``brainstorming" OR "revis*" OR ``feedback") \textbf{AND}
(``interface" OR ``system" OR ''prototype" OR ``tool") \textbf{AND} 
(``creative writing" OR ``story" OR ``fiction" OR ``narrative" OR ``poe*" OR ``script")
\end{quote}
In an effort to be inclusive, we expanded our search to all eligible works in the ACM Digital Library, ACL Anthology, and Web of Science (a multidisciplinary database including IEEE, Springer Nature, Elsevier, and more). Our search process returned 287 papers from all combined sources, excluding duplicates.
We screened paper keywords, titles, and abstracts against the eligibility criteria (\autoref{tab:inclusion-criteria}), resulting in 170 entries. One author read the papers fully to assess eligibility, which resulted in a final corpus of 67 papers (see the process in the flow graph \autoref{fig:lit-flow}).

\subsubsection{Secondary Literature Search}

Based on the initial findings, we conducted a secondary literature search to ensure comprehensive coverage and validate our findings.
Guided by the sub-themes of writers' needs that emerged from our analysis (see \autoref{tab:concise-theme}), we searched for literature that addresses the needs not covered in our initial search. We broadened our original search terms and included additional search terms (i.e. ideation, pre-writing, spark, organization) to obtain 468 papers, after which we filtered out those initially surveyed. This was followed by removing papers that did not fulfill inclusion criteria I, II, and III from \autoref{tab:inclusion-criteria}. 
We found an additional 11 papers, which we discuss in our results, thereby refining the analysis, adding nuance, and ensuring the robustness of our conclusions. We include a table listing all the references from the initial and secondary search in \autoref{appendix:lit-tools-full}.

\begin{table}
    \centering
    \caption{Demographic distribution of our participant pool for the online questionnaire.}
\begin{tabular}{@{}llc@{}}
\toprule
\textbf{Label}                    & \textbf{Values}         & \textbf{Count} \\ \midrule
\multirow{4}{*}{Age}     & 18-24          & 10    \\
                         & 25-34          & 7     \\
                         & 35-44          & 4     \\
                         & 45-54          & 1     \\ \midrule
\multirow{4}{*}{Gender} & Female         & 12    \\
                         & Male           & 6     \\
                         & Non-Binary     & 2     \\
                         & Prefer not to disclose &  2 \\ \midrule
\multirow{2}{*}{Genre}  & Fiction        & 20    \\
                         & Theatre & 2 \\ \midrule
\multirow{3}{*}{Country} & Canada         & 18    \\
                         & United States  & 2     \\
                         & United Kingdom & 1     \\ \bottomrule
\end{tabular}
    \label{tab:questionnaire-demographic}
\end{table}

\subsection[]{Online Questionnaire with Writers \includegraphics[height=0.7\baselineskip]{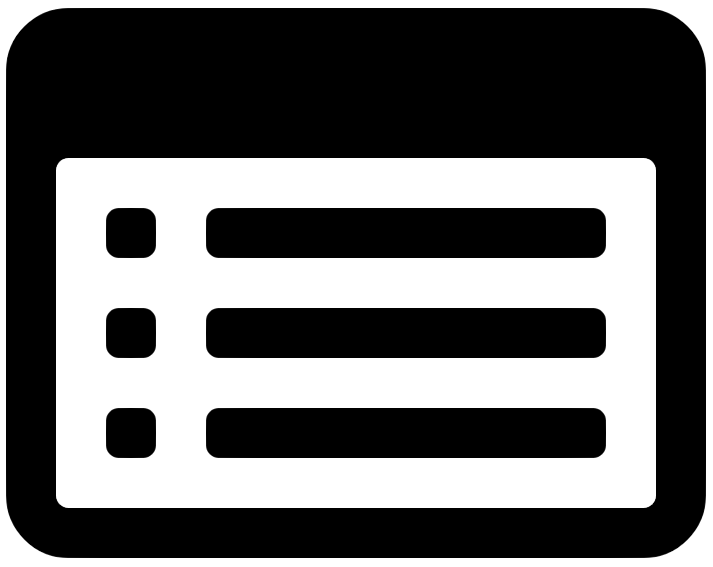}}
We deployed an online questionnaire to gather responses from practicing writers about their writing workflow, practices, and digital tools to facilitate writing tasks. 
Like Gero et al.~\cite{gero-2023-social}, we define writers as anyone self-identified as a creative writer, thereby being inclusive of those who engage in activities to produce creative writing artifacts beyond professional writers only. 
We recruited writers through our university's creative writing community mailing list, word of mouth, and online writing/creative communities on social media (e.g. X formerly Twitter, Reddit, Discord). 
We initially advertised the study publicly over X and received more than 400 responses. After filtering out responses with incomplete answers and AI generated text (using GPTZero\footnote{https://gptzero.me/}), we were left with 4 responses that we manually inspected to ensure their quality. 
Afterward, we limited recruitment to private Discord writing groups with user verifications.
Our questionnaire (available in the supplemental) was administered using Qualtrics\footnote{https://www.qualtrics.com/} and took around 10-15 minutes. Participants were compensated \$10. In total, we collected 22 complete responses and summarize participant demographic in  \autoref{tab:questionnaire-demographic}. Participants all had prior creative writing experience in either fiction or theatre plays (\(\mu\) = 9.5 years) and were located in Canada, the USA, or the UK.

\begin{figure*}[t]
    \centering
    \includegraphics[width=\linewidth]{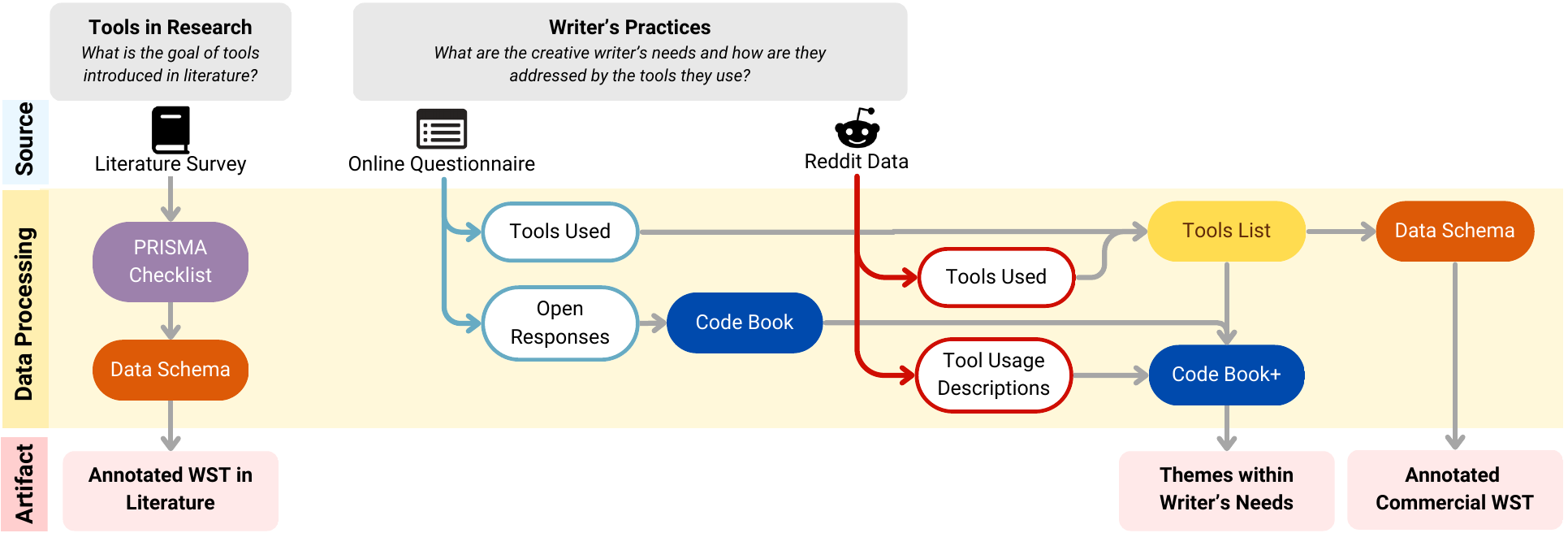}
    \caption{Methodology for processing data to be used in our analysis. We first collected \textit{(top)} data from three sources, and conducted \textit{(middle)} different annotation processes to obtain \textit{(bottom)} three data artifacts.}
    \label{fig:data-processing}
    \Description[Data processing flow diagram]{
    Three parallel flow diagram showing the methodology for processing data separated into three main sections, source, data processing, and the artifacts generated. Three sources were chosen including literature, online questionnaire, and Reddit data.
    }
\end{figure*}

\subsection[]{Reddit Data \includegraphics[height=1.2\fontcharht\font`\T]{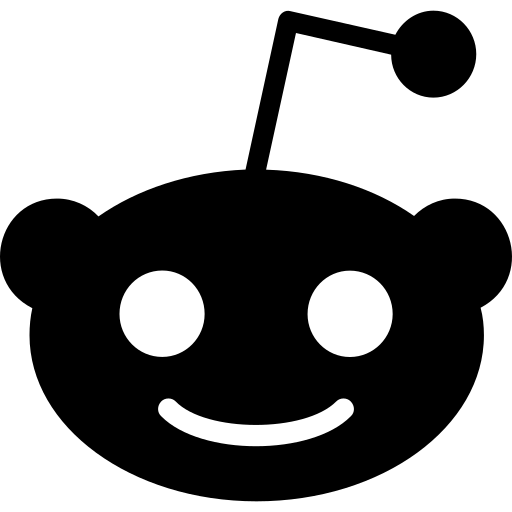}}
To gather more insight into the tools writers use and why they use them, we analysed Reddit discussion thread. We chose \textit{r/Writing}, a subreddit for practicing and aspiring writers to discuss writing-related topics, for its high subscriber count (2.9M members) and frequent activity (>100 users online on average). We extracted the data from Watchful1~\cite{reddit} between January 1, 2019 to December 31, 2023. Reddit data is separated into two components: submissions and comments to the submissions. We gathered the titles and body of submissions that contained the word \textit{``tool''} and manually inspected each matching entry to filter out irrelevant submissions. We included posts and comments for 357 submissions, resulting in a dataset with 4079 rows where the median length was 221 characters.


\section{Data Processing}
\label{sec:processing}


\subsection{Analyzing Literature}
We first reviewed literature from our initial search for insights into creative writing needs and how tools from prior research aimed to meet those needs. We build off existing design spaces~\cite{lee_design_2024} and taxonomies~\cite{frich_mapping_2019} to develop a data schema (\autoref{sec:final_schema}) used to annotate the literature. An overview of the process is shown in \autoref{fig:data-processing}.

\subsubsection{Annotation Process}
\label{sec:screening}
We created a preliminary schema by combining the design space for intelligent WST by Lee et al.~\cite{lee_design_2024} and the CST taxonomy by Frich et al.~\cite{frich_mapping_2019}. One coder conducted the initial coding using the preliminary schema, followed by rounds of calibration to solidify the data schema and ensure consistency. During calibration, we picked 20 random works (25\%) and separated them into two sets of 10. Four additional coders were separated into two groups, and one of the two sets was assigned to code independently. Afterwards, all coders met to discuss and compare the coding to assess the data schema. Code discrepancies were resolved through verbal ``negotiation agreement”~\cite{garrison2006revisiting}. One coder annotated the full corpus after finalizing the data schema (\autoref{sec:final_schema}).

\subsubsection{Data Schema for WST}
\label{sec:final_schema}
Our final data schema comprises nine dimensions: 
target population\footnote{\label{newcode} Newly introduced code.}, object of interest\footnoteref{newcode}, 
part of the writing process\footnote{Builds on the cognitive process model of writing model~\cite{flower_cognitive_1981}.}, 
writing context\footnote{Borrowed from Lee et al.~\cite{lee_design_2024}}, 
device\footnote{\label{frich} Borrowed from Frich et al.~\cite{frich_mapping_2019}}, 
maturity\footnoteref{frich}, 
tool complexity\footnoteref{frich}, 
evaluation\footnoteref{frich}, 
and collaboration paradigm\footnote{Expanded from Frich et al. with AI collaboration paradigms defined by Morris et al.~\cite{morris2023levels}}. Target population codes the tool's target user base, which includes children, experts (e.g. professional novelists), and novices. The object of interest is used to capture whether the tool's goal is to support the creation of the creative text itself or abstract components of the text (e.g. character, plot, setting, etc.). 
The code scheme is available in Appendix \ref{appendix:data-schema}, and our final categorized literature summarized in \autoref{fig:lit-tools} with details available in Appendix \ref{appendix:lit-tools-full}.

\subsection{Thematic Analysis of the Writing Process}
We used thematic analysis of responses from our questionnaire and Reddit data to identify the needs of current practicing writers. An overview of the process is shown in \autoref{fig:data-processing}. We split the data into subsets relevant to the 50 most mentioned tools across both data sources. Following, we compiled all our text data and performed a broad greedy tag for each text segment by asking \textbf{when} and \textbf{why} if the writer was using this tool.
A researcher assigned codes and extracted representative quotes, and then three authors passed through the codes and quotes to ensure agreement over the tagging. First, we identified recurring trends with online questionnaire responses and categorized them into different sub-themes. Then, we used the sub-themes discovered in the questionnaires to aid the exploration and categorization of Reddit data. We present an overview of the themes and sub-themes in \autoref{tab:concise-theme} and discuss details in the next sections.

\subsection{Collecting and Annotating Public WST}
We surveyed tools used by writers through our online questionnaire and Reddit data to investigate what writers use and how. The questionnaire prompted the writers to share tools they previously tried, why they liked or disliked the tool, and their purpose for using the tool. For the Reddit data, we used two procedures in parallel and merged the tools and information obtained from both, see \autoref{fig:data-processing}. Within \textit{r/Writing}, there is a weekly discussion post for WST called ``Writing Tools, Software, and Hardware'' which we used to obtain a list of tools and how writers use them. Concurrently, we preprocessed our Reddit dataset and used \texttt{spacy}'s named entity recognizer (NER) pipeline to extract a list of potential WST. We merged the tools mentioned in the questionnaire and those on Reddit to create a dataset containing 111 commercial tools. We annotated each tool using the same data schema as the literature review (see~\autoref{sec:final_schema}) and ranked the tools based on how often it was mentioned.

The remainder of the paper discusses insights resulting from our analysis. We structure results by main findings and provide takeaway insights, presenting opportunities for future researchers to explore. We use iconography to refer to the source of the data from which statements are drawn: literature
\aptLtoX{\includegraphics[height=1.2\fontcharht\font`\T]{img/lit-filled.png}}{\textsuperscript{\includegraphics[height=1.2\fontcharht\font`\T]{img/lit-filled.png}}}, 
online questionnaire
\aptLtoX{\includegraphics[height=1.2\fontcharht\font`\T]{img/ques-filled.png}}{\textsuperscript{\includegraphics[height=1.2\fontcharht\font`\T]{img/ques-filled.png}}}, or Reddit
\aptLtoX{\includegraphics[height=1.2\fontcharht\font`\T]{img/reddit-filled.png}}{\textsuperscript{\includegraphics[height=1.2\fontcharht\font`\T]{img/reddit-filled.png}}}\footnote{Following best practices~\cite{fiesler2024remember}, we directly quote Reddit users who provided explicit consent to be quoted and identified, whereas quotes from those who did not respond to our request are paraphrased and labelled with \protect\includegraphics[height=1.2\fontcharht\font`\T]{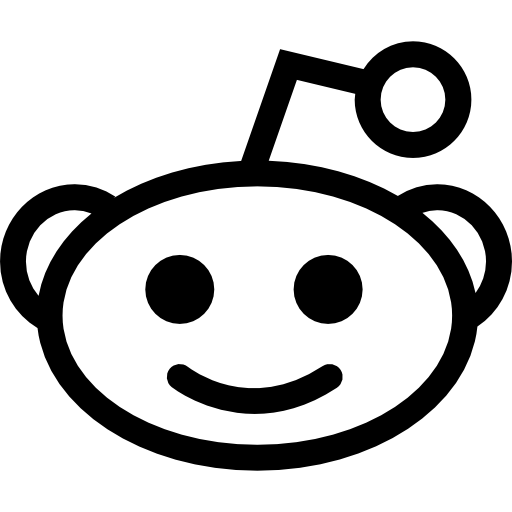}{{\small anon}}}. We also provide interactive visualizations of our data artifacts on a public-facing site\footnote{https://thewriteconnection.github.io/} and in \autoref{appendix:commercial-tools-full}.



\section{The writing process: it's more than we may think}

The research on WST\aptLtoX{\includegraphics[height=1.2\fontcharht\font`\T]{img/lit-filled.png}}{\textsuperscript{\includegraphics[height=1.2\fontcharht\font`\T]{img/lit-filled.png}}} is heavily focused on and grounded in the cognitive process model of writing~\cite{flower_cognitive_1981}, which identifies four main processes: planning, generating, revising, and monitoring. Planning involves ideation, organization, and goal setting for the work. Generating refers to translating the ideas into words. Revising consists of reflecting, reading, and editing. These processes are monitored closely against each other continuously, non-linearly.
This model has helped guide and evaluate the development of tools that support these processes. For example, most of the commercial tools and literature focus specifically on planning~(e.g. ~\cite{bala_2022,chou_2023}), generating~(e.g. ~\cite{yuan_wordcraft_2022,dang_choice_2023,mirowski_co-writing_2023}) or revising~(e.g. ~\cite{gibson_reflective_2017,shum_reflecting_2016}), as shown in \autoref{fig:lit-tools}.

Although the model provides researchers and developers with a scope, taking the model too prescriptively might have crystallized an isolated view of each process. We break down examples of tools and how they may perpetuate the isolated view in \S\ref{sec:writing-activity}. Furthermore, this model only captures processes involved while \emph{actively writing} with a clearly defined goal and direction; whereas, we found in our questionnaire\aptLtoX{\includegraphics[height=1.2\fontcharht\font`\T]{img/ques-filled.png}}{\textsuperscript{\includegraphics[height=1.2\fontcharht\font`\T]{img/ques-filled.png}}}
and Reddit\aptLtoX{\includegraphics[height=1.2\fontcharht\font`\T]{img/reddit-filled.png}}{\textsuperscript{\includegraphics[height=1.2\fontcharht\font`\T]{img/reddit-filled.png}}} data 
the writers have much more to their writing process than the writing task itself. A large component of the creative writing process consists of an extensive \emph{pre-writing} stage. 
For many writers, the pre-writing process begins long before plotting or drafting. Gathering inspiration can take years. For example, inspiration for writers such as J.R.R. Tolkein~(\textit{Lord of the Rings}) and Han Kang~(\textit{The Vegetarian}) originated from quotes they came across years before starting their first drafts~\cite{Tolkein_1984,Shin_2016}. The term pre-writing was first popularized by Rohman~\cite{rohman1965pre} as another term for what other writing models call planning. However, writers have now adopted the term more colloquially to refer to the process before any writing activity.

There are various models of the creative process at large, the most famous ones being Wallas's~\cite{wallas1926art} and an extension of it by Csikszentmihalyi~\cite{csikszentmihalyi1997flow}. Both models define several phases: preparation, incubation, illumination~(or insight), verification~(or evaluation), and elaboration~(unique to Csikszentmihalyi). Preparation refers to a period when the individual becomes immersed, consciously or not, within a topic and acquires new knowledge. Incubation is when ideas churn unconsciously and form connections with each other. Then is illumination, or insight, when that \textit{``Eureka''} moment strikes and the idea emerges to the conscious level. Verification, or evaluation, is where the individual evaluates if the idea is worth pursuing. Then elaboration, not defined in Wallas's model, is when the individual works to externalize their ideas into a creative work. To characterize creative writing as both a creative process and a writing process, we align the theoretical models for creativity~\cite{wallas1926art,csikszentmihalyi1997flow}, the cognitive model of the writing process~\cite{flower_cognitive_1981}, and the six identified parts of the creative process that CSTs address~\cite{frich_mapping_2019} in Table \ref{tab:writing-process}. We propose pre-writing to encompass the aspects of creative writing which are associated with the four phases within Wallas's creative process model: acquiring knowledge~(\textit{preparation}), internalizing the new material~(\textit{incubating}), generating new ideas~(\textit{illumination}), and evaluating those new ideas~(\textit{verification}). 

By grounding our definition of pre-writing as a creative process, we hope to shed light on this stage which has been overlooked due to the lack of definition for this process. The following sections first go deeper into the phases within the pre-writing stage, focusing on the writers’ unmet needs. We then present our findings on how prior research and current commercial WST support the cognitive processes associated with writing before launching into our discussion. 

\begin{table*}[]
    \centering
    \caption{Aspects of the creative writing process identified by previous surveys on CSTs~\cite{frich_mapping_2019} and WSTs~\cite{lee_design_2024,gero_design_2022}, alongside the cognitive writing model~\cite{flower_cognitive_1981}, and theoretical models of creativity~\cite{wallas1926art,csikszentmihalyi1997flow}.}
    \label{tab:writing-process}
    \begin{tabular}
    {p{0.1\linewidth}p{0.1\linewidth}p{0.18\linewidth}p{0.2\linewidth}p{0.16\linewidth}}
\toprule
\textbf{Ours} & \textbf{Wallas~\cite{wallas1926art}} & \textbf{Csikszentmihalyi~\cite{csikszentmihalyi1997flow}} & \textbf{Flower and Hayes~\cite{flower_cognitive_1981}} & \textbf{Frich et al.~\cite{frich_mapping_2019}} \\ \midrule
\textit{\textbf{Pre-writing}} & Preparation & Preparation &  & Pre-ideation \\
\textit{\textbf{}} & Incubation & Incubation &  &  \\
\textit{\textbf{}} & Illumination & Insight & Planning & Ideation \\
\textit{\textbf{}} & Verification & Evaluation &  & Evaluation \\
\textit{\textbf{Planning}} &  & Elaboration &  & Implementation \\
\textit{\textbf{Generating}} &  &  & Generating &  \\
\textit{\textbf{Revising}} &  &  & Revising & Iteration \\
\textit{\textbf{Monitoring}} &  &  & Monitoring & Project Management \\
\bottomrule
\end{tabular}

\end{table*}


\setlength{\fboxrule}{0.5pt}
\aptLtoX{\begin{table*}[t]
\small
\caption[]{Themes and sub-themes found from the qualitative analysis of our data sources. In columns \includegraphics[height=6px]{img/reddit-filled.png} and \includegraphics[height=6px]{img/ques-filled.png}, we label the presence of evidence from each source, with the value ranging from no evidence (\squarelight), one to a few instances of anecdotal evidence (\squaremedium), to numerous instances (\squaredark). In column \includegraphics[height=1.2\fontcharht\font`\T]{img/lit-filled.png}, we list a single example of prior work that addresses each sub-theme, those labeled \textit{None} had no works in our literature survey that developed tools to address them. The TA column refers to the take-away derived from the themes. The last column is the row ID, to allow for cross-referencing.}
\begin{tabular}
{p{0.01\linewidth}p{0.09\linewidth}p{0.168\linewidth}p{0.47\linewidth}p{0.01\linewidth}p{0.01\linewidth}p{0.025\linewidth}p{0.005\linewidth}p{0.005\linewidth}}
& \textbf{Theme}              & \textbf{Sub-theme}      & \textbf{Description}                                            & \includegraphics[height=1.2\fontcharht\font`\T]{img/reddit-filled.png}  &  \includegraphics[height=1.2\fontcharht\font`\T]{img/ques-filled.png}                              & \includegraphics[height=1.2\fontcharht\font`\T]{img/lit-filled.png}  & \textbf{TA}             & \textbf{ID} \\
\hline
\multirow{6}{*}{{\textbf{Pre-writing}}}
& \cellcolor[HTML]{FFEABE}Creativity   & Community  & Writers look to form a community of practice.                             & \squaredark & \squarelight & \cite{huang_heteroglossia_2020} & 1  & \textit{a} \\
 &  \cellcolor[HTML]{FFEABE} & Unguided exploration   & Writers use tools that let them passively explore without a goal.       & \squaredark & \squarelight & \cite{sadauskas_mining_2015} & 2  & \textit{b} \\
 &  \cellcolor[HTML]{FFEABE} & Evaluating inspiration & Writers evaluate ideas for potential.                                   & \squaremedium & \squaremedium & \cite{chou_2023} & 2  & \textit{c} \\
 & \cellcolor[HTML]{D5F5C6}Recollection      & Resources & Writers look for guides and resources from other writers. & \squaredark & \squarelight & \cite{luck_creative_2012} & 1  & \textit{d} \\
 & \cellcolor[HTML]{D5F5C6} & Capture spark & Writers need tools to capture their creative sparks.             & \squaredark & \squaredark & \textit{None} & 2  & \textit{e} \\
 & \cellcolor[HTML]{E5D9ED}Organization & Organizing inspiration & Writers organize their ideas and inspirations.               & \squaredark & \squaremedium & \cite{chi_2010} & 2  & \textit{f} \\ 
 \hline
\multirow{4}{*}{\textbf{Planning}}
& \cellcolor[HTML]{FFEABE}Creativity   & Ideating with tools    & Writers may use tools to generate ideas.                    & \squaredark & \squaremedium & \cite{settles-2010-computational} & 3  & \textit{g} \\
 &  \cellcolor[HTML]{FFEABE} & Developing plans       & Writers use templates and guides during initial planning.             & \squaremedium & \squaredark & \cite{ashida_plot-creation_2019} & 5  & \textit{h} \\
 & \cellcolor[HTML]{D5F5C6}Recollection  & Visualizing elements   & Writers use visual representation for elements in their story.  & \squaremedium & \squaredark & \cite{hoque_dramatvis_2022} & 4  & \textit{i} \\
 & \cellcolor[HTML]{E5D9ED}Organization & Organizing elements    & Writers use tools that help them organize elements.        & \squaredark & \squaredark & \cite{xu_jamplate_2024} & 5  & \textit{j} \\  
 \hline
\multirow{4}{*}{{\textbf{Generating}}}
 & \cellcolor[HTML]{FFEABE}Creativity   & Non-linear writing     & Writers write sections, like scenes, out of order.         & \squaremedium & \squaredark & \cite{linaza2004authoring} & 6  & \textit{k} \\
& \cellcolor[HTML]{FFEABE} & AI support  & Writers have varying opinions on using AI during generation.                        & \squaredark & \squaredark & \cite{dang_choice_2023} &    & \textit{l} \\
 & \cellcolor[HTML]{FFEABE} & Writing environment    & Writers spend time to discover their own workflow.                     & \squaredark & \squarelight & \cite{goncalves_olfactory_2017} & 7  & \textit{m} \\
 & \cellcolor[HTML]{E5D9ED}Organization & Ease of access         & Writers prefer tools that are easily accessible across devices. & \squaredark & \squaremedium & \cite{bonsignore_2013} & 7  & \textit{n} \\  
 \hline
\multirow{6}{*}{{\textbf{Revising}}}
 & \cellcolor[HTML]{FFEABE}Creativity & Editing with others    & Writers seek out help of others to revise their work.        & \squaremedium & \squaredark & \cite{swanson_say_2008} & 8  & \textit{o} \\
 &  \cellcolor[HTML]{FFEABE} & Alternative views      & Writers transform their text into alternative versions to stimulate creativity and identify errors. & \squaredark & \squaredark & \cite{booten_poetry_2021} & 9  & \textit{p} \\
 & \cellcolor[HTML]{D5F5C6}Recollection & Editing with machines  & Writers want tools that give them specific feedback.                  & \squaredark & \squaremedium & \cite{sanghrajka_lisa_2017} & 8  & \textit{q} \\
 & \cellcolor[HTML]{E5D9ED}Organization       & Structural revision    & Writers make large scale revisions involving shifting large sections of texts. & \squaremedium & \squaredark & \cite{trigg_hypertext_1987} & 9  & \textit{r} \\  
 \hline
\multirow{3}{*}{{\textbf{Monitoring}}}
 & \cellcolor[HTML]{D5F5C6}Recollection &  Peripheral reminders   & Writers keep information relevant to text in visible locations while writing.       & \squaremedium & \squaredark & \cite{singh2023hide} & 10 & \textit{s} \\
 & \cellcolor[HTML]{E5D9ED}Organization & Support for projects   & Writers want more support from tools that help them organize large projects. & \squaredark & \squaredark & \cite{holdich_improving_2004} & 10 & \textit{t}\\
\end{tabular}
\label{tab:concise-theme}
\end{table*}}{
\begin{table*}[t]
\small
\caption{Themes and sub-themes found from the qualitative analysis of our data sources. In columns \includegraphics[height=6px]{img/reddit-filled.png} and \includegraphics[height=6px]{img/ques-filled.png}, we label the presence of evidence from each source, with the value ranging from no evidence (\squarelight), one to a few instances of anecdotal evidence (\squaremedium), to numerous instances (\squaredark). In column \includegraphics[height=6px]{img/lit-filled.png}, we list a single example of prior work that addresses each sub-theme, those labeled \textit{None} had no works in our literature survey that developed tools to address them. The TA column refers to the take-away derived from the themes. The last column is the row ID, to allow for cross-referencing.}
\begin{tabular}
{p{0.01\linewidth}p{0.09\linewidth}p{0.168\linewidth}p{0.47\linewidth}p{0.01\linewidth}p{0.01\linewidth}p{0.025\linewidth}p{0.005\linewidth}p{0.005\linewidth}}
& \textbf{Theme}              & \textbf{Sub-theme}      & \textbf{Description}                                            & \includegraphics[height=1.2\fontcharht\font`\T]{img/reddit-filled.png}  &  \includegraphics[height=1.2\fontcharht\font`\T]{img/ques-filled.png}                              &\includegraphics[height=1.2\fontcharht\font`\T]{img/lit-filled.png}  & \textbf{TA}             & \textbf{ID} \\
\thickhline
\parbox[t]{2mm}{\centering\multirow{6}{*}{\rotatebox[origin=c]{90}{\textbf{Pre-writing}}}}
& \cellcolor[HTML]{FFEABE}Creativity   & Community  & Writers look to form a community of practice.                             & \squaredark & \squarelight & \cite{huang_heteroglossia_2020} & 1  & \textit{a} \\
 &  \cellcolor[HTML]{FFEABE} & Unguided exploration   & Writers use tools that let them passively explore without a goal.       & \squaredark & \squarelight & \cite{sadauskas_mining_2015} & 2  & \textit{b} \\
 &  \cellcolor[HTML]{FFEABE} & Evaluating inspiration & Writers evaluate ideas for potential.                                   & \squaremedium & \squaremedium & \cite{chou_2023} & 2  & \textit{c} \\
 & \cellcolor[HTML]{D5F5C6}Recollection      & Resources & Writers look for guides and resources from other writers. & \squaredark & \squarelight & \cite{luck_creative_2012} & 1  & \textit{d} \\
 & \cellcolor[HTML]{D5F5C6} & Capture spark & Writers need tools to capture their creative sparks.             & \squaredark & \squaredark & \textit{None} & 2  & \textit{e} \\
 & \cellcolor[HTML]{E5D9ED}Organization & Organizing inspiration & Writers organize their ideas and inspirations.               & \squaredark & \squaremedium & \cite{chi_2010} & 2  & \textit{f} \\ 
 \hline
\parbox[t]{2mm}{\centering\multirow{4}{*}{\rotatebox[origin=c]{90}{\textbf{Planning}}}}
& \cellcolor[HTML]{FFEABE}Creativity   & Ideating with tools    & Writers may use tools to generate ideas.                    & \squaredark & \squaremedium & \cite{settles-2010-computational} & 3  & \textit{g} \\
 &  \cellcolor[HTML]{FFEABE} & Developing plans       & Writers use templates and guides during initial planning.             & \squaremedium & \squaredark & \cite{ashida_plot-creation_2019} & 5  & \textit{h} \\
 & \cellcolor[HTML]{D5F5C6}Recollection  & Visualizing elements   & Writers use visual representation for elements in their story.  & \squaremedium & \squaredark & \cite{hoque_dramatvis_2022} & 4  & \textit{i} \\
 & \cellcolor[HTML]{E5D9ED}Organization & Organizing elements    & Writers use tools that help them organize elements.        & \squaredark & \squaredark & \cite{xu_jamplate_2024} & 5  & \textit{j} \\  
 \hline
 \parbox[t]{2mm}{\centering\multirow{4}{*}{\rotatebox[origin=c]{90}{\textbf{Generating}}}}
 & \cellcolor[HTML]{FFEABE}Creativity   & Non-linear writing     & Writers write sections, like scenes, out of order.         & \squaremedium & \squaredark & \cite{linaza2004authoring} & 6  & \textit{k} \\
& \cellcolor[HTML]{FFEABE} & AI support  & Writers have varying opinions on using AI during generation.                        & \squaredark & \squaredark & \cite{dang_choice_2023} &    & \textit{l} \\
 & \cellcolor[HTML]{FFEABE} & Writing environment    & Writers spend time to discover their own workflow.                     & \squaredark & \squarelight & \cite{goncalves_olfactory_2017} & 7  & \textit{m} \\
 & \cellcolor[HTML]{E5D9ED}Organization & Ease of access         & Writers prefer tools that are easily accessible across devices. & \squaredark & \squaremedium & \cite{bonsignore_2013} & 7  & \textit{n} \\  
 \hline
\parbox[t]{2mm}{\centering\multirow{5}{*}{\rotatebox[origin=c]{90}{\textbf{Revising}}}}
 & \cellcolor[HTML]{FFEABE}Creativity & Editing with others    & Writers seek out help of others to revise their work.        & \squaremedium & \squaredark & \cite{swanson_say_2008} & 8  & \textit{o} \\
 &  \cellcolor[HTML]{FFEABE} & Alternative views      & Writers transform their text into alternative versions to stimulate creativity and identify errors. & \squaredark & \squaredark & \cite{booten_poetry_2021} & 9  & \textit{p} \\
 & \cellcolor[HTML]{D5F5C6}Recollection & Editing with machines  & Writers want tools that give them specific feedback.                  & \squaredark & \squaremedium & \cite{sanghrajka_lisa_2017} & 8  & \textit{q} \\
 & \cellcolor[HTML]{E5D9ED}Organization       & Structural revision    & Writers make large scale revisions involving shifting large sections of texts. & \squaremedium & \squaredark & \cite{trigg_hypertext_1987} & 9  & \textit{r} \\  
 \hline
 \parbox[t]{2mm}{\centering\multirow{3}{*}{\rotatebox[origin=c]{90}{\textbf{Monitoring}}}}
 & \cellcolor[HTML]{D5F5C6}Recollection &  Peripheral reminders   & Writers keep information relevant to text in visible locations while writing.       & \squaremedium & \squaredark & \cite{singh2023hide} & 10 & \textit{s} \\
 & \cellcolor[HTML]{E5D9ED}Organization & Support for projects   & Writers want more support from tools that help them organize large projects. & \squaredark & \squaredark & \cite{holdich_improving_2004} & 10 & \textit{t}\\
\end{tabular}
\label{tab:concise-theme}
\end{table*}}
\setlength{\fboxrule}{3pt}
\renewcommand\fbox{\fcolorbox{gray!15}{gray!15}}


\section{A Focus on Pre-writing: An Overlooked Stage}
\label{sec:6}

Flower and Hayes~\cite{flower_cognitive_1981} define the start of writing activity as when the writer formulates a goal to achieve, involving an element that would be part of the written text. For narratives, it would include elements like plot, character, setting, and theme~\cite{bal2009narratology}. While these elements are central to narratives, the writer's inspiration process has not received much attention within WST research. Through our empirical analysis of the writer’s process \aptLtoX{\includegraphics[height=1.2\fontcharht\font`\T]{img/ques-filled.png}\includegraphics[height=1.2\fontcharht\font`\T]{img/reddit-filled.png}}{\textsuperscript{\includegraphics[height=1.2\fontcharht\font`\T]{img/ques-filled.png}\includegraphics[height=1.2\fontcharht\font`\T]{img/reddit-filled.png}}}, we found that these elements are the product of a pre-writing stage. The phases that we observed within pre-writing closely matched the Wallas creative process model~\cite{wallas1926art} and the first four phases of Csikszentmihalyi's creative flow process~\cite{csikszentmihalyi1997flow}. We map the four phases of the creative model to several sub-themes found within the writer’s process. Overall, we find that the research space surrounding pre-writing is vast and uncharted.

\subsection{Seeking Inspiration}
\label{sec:seek_inspo}

Writers look for sources of inspiration by learning from more experienced writers, browsing the internet, reading books, chatting with friends, and through personal experiences. 
We found over 500 submissions and over 15k comments on Reddit \aptLtoX{\includegraphics[height=1.2\fontcharht\font`\T]{img/ques-filled.png}}{\textsuperscript{\includegraphics[height=1.2\fontcharht\font`\T]{img/reddit-filled.png}}} discussing different methods of finding sources of inspiration. 

\subsubsection{Resources and guides (\autoref{tab:concise-theme}, row d)}
Novice writers often seek insights into \textit{``what makes good writing great''}~(\includegraphics[height=1.2\fontcharht\font`\T]{img/reddit-logo.png}{anon}) by asking about well-known authors' routines, methods, and inspirations. 
Experienced writers sometimes consolidate their own experiences to create guides for other writers; some are shared on Reddit threads or personal blogs. 
We observed that many writers on Reddit ask for guides to sources of inspiration and receive a wide range of opinions and methods.
For example, some suggest \textit{``find[ing] a routine and writ[ing] often''}~(\includegraphics[height=1.2\fontcharht\font`\T]{img/reddit-logo.png}{anon}) as \textit{``a good writing routine makes it easier for inspiration to strike''}~(\includegraphics[height=1.2\fontcharht\font`\T]{img/reddit-logo.png}{anon}) while others suggest \textit{``look[ing] for new experiences''}~(\includegraphics[height=1.2\fontcharht\font`\T]{img/reddit-logo.png}{anon}) as \textit{``inspiration comes from seeing places, experiencing things, and wondering `oh, what if?'''}~(\includegraphics[height=1.2\fontcharht\font`\T]{img/reddit-logo.png}{anon}) 
These diverse opinions highlight novices' difficulty in determining a method that works for them, even if they encounter a guide. As we did not find prior work\aptLtoX{\includegraphics[height=1.2\fontcharht\font`\T]{img/lit-filled.png}}{\textsuperscript{\includegraphics[height=1.2\fontcharht\font`\T]{img/lit-filled.png}}} addressing this need, we searched for relevant literature on providing resources and guides for creative writers. We found one work identifying considerations for developing a creative-writing digital-learning and material-management system~\cite{luck_creative_2012}. So, although some research surrounds this topic, prior work has not proposed solutions through WST.

\subsubsection{Inspiration through unguided exploration (\autoref{tab:concise-theme}, row b)}
Writers may be inspired by something on social media, visuals, music, or personal experiences\aptLtoX{\includegraphics[height=1.2\fontcharht\font`\T]{img/reddit-filled.png}}{\textsuperscript{\includegraphics[height=1.2\fontcharht\font`\T]{img/reddit-filled.png}}}. Browsing image galleries: \textit{`` Pinterest for building collections of images/videos based on genres that inspire [them]''}~(\includegraphics[height=1.2\fontcharht\font`\T]{img/reddit-logo.png}{anon}), or collecting \textit{``random, abstract, famous quotes''}~(\includegraphics[height=1.2\fontcharht\font`\T]{img/reddit-logo.png}{anon}) to explore ideas passively, can stimulate inspiration. Auditory media are also leveraged. For instance, a writer mentioned that three of their current work's inspirations \textit{``came from music''} as they are \textit{``would imagine scenarios and lose [themselves] in creating different worlds and backstories''}~(\includegraphics[height=1.2\fontcharht\font`\T]{img/reddit-logo.png}{anon}).

By far, the most common source of inspiration is personal experience, with ideas originating, e.g. \textit{``when working in a music store selling guitars and pianos''}~(\includegraphics[height=1.2\fontcharht\font`\T]{img/reddit-logo.png}{anon}), or from \textit{``books [they] read as a kid''}~(\includegraphics[height=1.2\fontcharht\font`\T]{img/reddit-logo.png}{anon}).
Writers often passively brainstorm by reflecting on their and others' experiences. Inspiration can come from \textit{``real life and real people/situations''}~(\includegraphics[height=1.2\fontcharht\font`\T]{img/reddit-logo.png}{anon}) or \textit{``certain characters or establishments in other books they've read and enjoyed''}~(\includegraphics[height=1.2\fontcharht\font`\T]{img/reddit-logo.png}{anon}). 
Interestingly, there are also conversations on dreams individuals had as an \textit{``immense source of creativity''}~(\includegraphics[height=1.2\fontcharht\font`\T]{img/reddit-logo.png}{anon}). Using dreams as a source of inspiration has been explored by prior work in visual story creation~\cite{liu-2024}.
One way in which prior work\aptLtoX{\includegraphics[height=1.2\fontcharht\font`\T]{img/lit-filled.png}}{\textsuperscript{\includegraphics[height=1.2\fontcharht\font`\T]{img/lit-filled.png}}} has addressed writers' desires to explore their own experiences is through a tool for exploring one's social media to search for inspiration~\cite{sadauskas_mining_2015}. 
Other tangential work features assistance for songwriting by generating song titles that lyricists can use as inspiration~\cite{settles_computational_2010}.

\subsubsection{Community of practice (\autoref{tab:concise-theme}, row a)}
Writers \aptLtoX{\includegraphics[height=1.2\fontcharht\font`\T]{img/reddit-filled.png}}{\textsuperscript{\includegraphics[height=1.2\fontcharht\font`\T]{img/reddit-filled.png}}} also find inspiration through discussions with others, with some engaging in a \textit{``ramble to a writing friend when [they] get stuck''} for the friend to \textit{``nudge [them] in the right direction''}~(\includegraphics[height=1.2\fontcharht\font`\T]{img/reddit-logo.png}{anon}). 
When writers cannot find peers in their close circle, they may turn to social media and find others to exchange ideas with. Little work has explored collaborative ideation within the context of creative writing\aptLtoX{\includegraphics[height=1.2\fontcharht\font`\T]{img/lit-filled.png}}{\textsuperscript{\includegraphics[height=1.2\fontcharht\font`\T]{img/lit-filled.png}}}. For example, one implemented a new method for ideating with MTurk workers~\cite{huang_heteroglossia_2020}. 
Our secondary search also led to work that explored how LLMs can act as partners to collaborate with writers during pre-writing ideation~\cite{wan-2024-prewriting}, which relates to forming a community.

\subsubsection{Organizing inspiration (\autoref{tab:concise-theme}, row f)}
One disadvantage of unguided exploration is that retrieving an inspirational source after the fact\aptLtoX{\includegraphics[height=1.2\fontcharht\font`\T]{img/reddit-filled.png}}{\textsuperscript{\includegraphics[height=1.2\fontcharht\font`\T]{img/reddit-filled.png}}} becomes difficult. 
Many wish there were \textit{``better bookmarking tool[s]''} because it is \textit{``really beneficial to save their inspirations in one place and easily retrieve them later''}~(\includegraphics[height=1.2\fontcharht\font`\T]{img/reddit-logo.png}{anon}). Writers typically find that \textit{``things begin with an accumulation of unrelated notes, in which patterns eventually become apparent''}~(\includegraphics[height=1.2\fontcharht\font`\T]{img/reddit-logo.png}{anon}) and \textit{``a spark of inspiration strikes''}~(\includegraphics[height=1.2\fontcharht\font`\T]{img/reddit-logo.png}{anon}). 
Prior work\aptLtoX{\includegraphics[height=1.2\fontcharht\font`\T]{img/lit-filled.png}}{\textsuperscript{\includegraphics[height=1.2\fontcharht\font`\T]{img/lit-filled.png}}} like Chi and Lieberman~\cite{chi_2010}, Landry~\cite{landry_storytelling_2008}, and Wong and Lee~\cite{wong_creative_2011} explores how to help people weave images and clips captured from their daily lives into a compelling narrative structure.

\subsection{Capturing the Creative Spark \textit{(\autoref{tab:concise-theme}, row e)}}
\label{sec:seize_spark}
The creative spark fuels the formation of writing goals, so writers need to quickly capture those moments of inspiration and \textit{``write them down even if it's the middle of the night''}~(\includegraphics[height=1.2\fontcharht\font`\T]{img/reddit-logo.png}{anon}). When struck with a creative spark, writers\aptLtoX{\includegraphics[height=1.2\fontcharht\font`\T]{img/ques-filled.png}}{\textsuperscript{\includegraphics[height=1.2\fontcharht\font`\T]{img/ques-filled.png}}} said they reached for whatever was available to write down their ideas quickly like the \textit{``back of a crumpled receipt''}~(\textit{P10}), 
notebooks they carry around~(\textit{P13} and \includegraphics[height=1.2\fontcharht\font`\T]{img/reddit-logo.png}{anons}),
notepad-like apps on their phones~(P9 and \includegraphics[height=1.2\fontcharht\font`\T]{img/reddit-logo.png}{anons}), \textit{``voice notes''}~(\includegraphics[height=1.2\fontcharht\font`\T]{img/reddit-logo.png}{anon}), or they \textit{``take a picture''}~(\includegraphics[height=1.2\fontcharht\font`\T]{img/reddit-logo.png}{anon}) to capture the moment. Some writers use idea notebooks to write down their ideas to sort through later. One writer, for instance, said that they \textit{``open up that journal and go through [their] undeveloped ideas and try to flesh them out in a short story''}~(\includegraphics[height=1.2\fontcharht\font`\T]{img/reddit-logo.png}{anon}).
However, many writers lament that they may forget or lose the notes of their ideas. We found no work within our literature review that evaluates tools that allow writers to capture their ideas easily;
no additional work was found in our secondary search either.

\subsection{Evaluating the Spark \textit{(\autoref{tab:concise-theme}, row c)}}
\label{sec:spark_worth}
Writers need to evaluate their spark to see whether it is worth translating into text. 
Sometimes the spark is not worth exploring, and 
\textit{``turns out to be this rubbish 3 second analog between me and a talking cardboard box''}~(\includegraphics[height=1.2\fontcharht\font`\T]{img/reddit-logo.png}{anon}).
We found no literature \aptLtoX{\includegraphics[height=1.2\fontcharht\font`\T]{img/lit-filled.png}}{\textsuperscript{\includegraphics[height=1.2\fontcharht\font`\T]{img/lit-filled.png}}} related to the evaluation process for rough, initial ideas in our corpus.
On the other hand, we found over 18k posts\aptLtoX{\includegraphics[height=1.2\fontcharht\font`\T]{img/reddit-filled.png}}{\textsuperscript{\includegraphics[height=1.2\fontcharht\font`\T]{img/reddit-filled.png}}} and 20 participants\aptLtoX{\includegraphics[height=1.2\fontcharht\font`\T]{img/ques-filled.png}}{\textsuperscript{\includegraphics[height=1.2\fontcharht\font`\T]{img/ques-filled.png}}} discussing different processes for evaluating writers' initial ideas, most of which involved a mostly reflective mental process which varies based on individual preferences. For example, some writers\aptLtoX{\includegraphics[height=1.2\fontcharht\font`\T]{img/ques-filled.png}}{\textsuperscript{\includegraphics[height=1.2\fontcharht\font`\T]{img/ques-filled.png}}} may use pen and paper~(2 participants), while others may prefer a word processor~(5 participants) or a combination of the two~(15 participants) to evaluate the idea. Once the idea becomes entrenched in the writer’s mind and they feel it is worth elaborating on, they embark on the writing activity, where most of the WST research centers. At this point, the idea has transformed into an element of the text. 
Although there is no prior work\aptLtoX{\includegraphics[height=1.2\fontcharht\font`\T]{img/lit-filled.png}}{\textsuperscript{\includegraphics[height=1.2\fontcharht\font`\T]{img/lit-filled.png}}} which specifically supports evaluating a spark, there are systems where a feature supports the writer's idea evaluation process~(e.g. TaleStream~\cite{chou_2023}). In the next section, we discuss systems that support more goal-oriented brainstorming.

\aptLtoX{\begin{shaded}
\textbf{Takeaway insights:}
\begin{itemize}
\item[1.] Pre-writing and processes associated with it are not well defined in literature.
\item[2.] Tools that help facilitate recollection and organization of inspirations are lacking~(\S\ref{sec:seek_inspo},\S\ref{sec:seize_spark}, S\ref{sec:spark_worth}).
\end{itemize}
\end{shaded}}{
\vspace{6px}
\noindent\fbox{%
\parbox{0.95\linewidth}{
\textbf{Takeaway insights:}
\begin{itemize}
\item[1.] Pre-writing and processes associated with it are not well defined in literature.
\item[2.] Tools that help facilitate recollection and organization of inspirations are lacking~(\S\ref{sec:seek_inspo},\S\ref{sec:seize_spark}, S\ref{sec:spark_worth}).
\end{itemize}
}}}

\section{Writing Activity: Where the Spotlight has Been}
\label{sec:writing-activity}
After pre-writing, we now step into the final phase of Csikszentmihalyi's~\cite{csikszentmihalyi1997flow} creative model: elaboration. As shown in \autoref{tab:writing-process}, we aligned the cognitive process model of writing~\cite{flower_cognitive_1981} within the elaboration phase. In this section, we examine how commercial tools and research have addressed the writer’s needs within each process. Moreover, in contrast to research and commercial WSTs for pre-writing, the tools for the writing activity are well established and abundant.

\begin{figure*}
    \centering
    \includegraphics[width=0.95\linewidth]{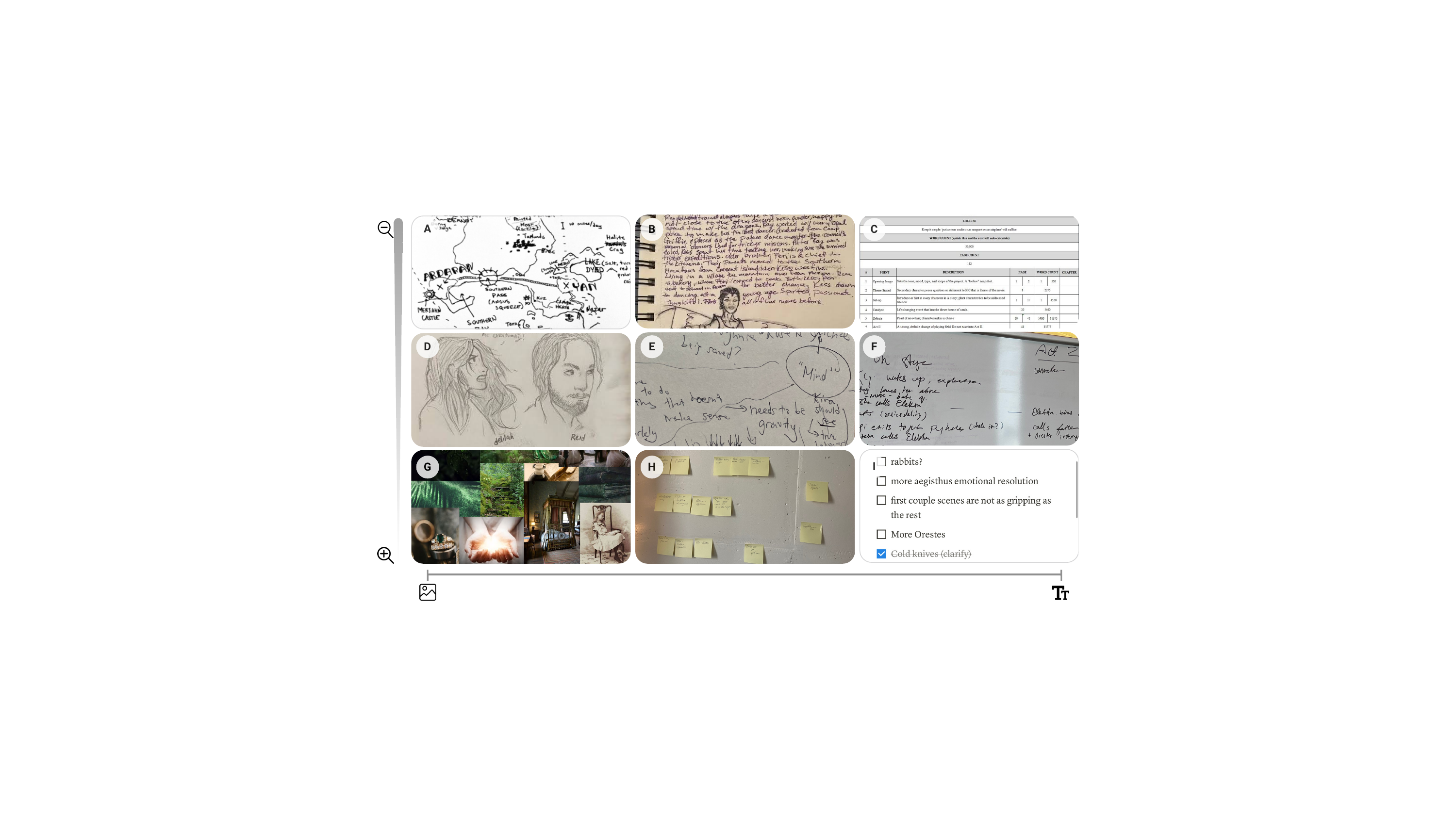}
    \caption{Artifacts shared by writers through our online survey representing different possibilities of representing text elements. The examples are arranged from image-based representations to text-based along the x-axis and at varying levels of specificity on the y-axis. Writers use representations like~(A) maps,~(B) character sheets,~(C) spreadsheets,~(D) character sketches,~(E) mind maps,~(F) act-level outlines,~(G) mood boards,~(H) spatially organized sticky notes and~(I) bullet point notes.}
    \Description[Writer visual aid artifacts]{Nine artifacts shared by writers to represent the variety in ways text can be represented by visuals. There is a drawing of a map, character sketches, sticky notes on the wall, whiteboard with text, mood board with images.}
    \label{fig:vis-examples}
\end{figure*}

\subsection{Planning: Handling Alternative Representations of Text}
\label{sec:planning}
During planning, the writer creates a representation of elements like plot points, characters, themes, etc.. Planning tools target needs such as brainstorming, guiding the planning process, representing story elements, and managing the writing task. Despite the number of commercial tools, research in this area is still lacking compared to the generation process.

\subsubsection{Brainstorming with tools (\autoref{tab:concise-theme}, row g)}
\label{sec:brainstorm}
During pre-writing, writers\aptLtoX{\includegraphics[height=1.2\fontcharht\font`\T]{img/reddit-filled.png}\includegraphics[height=1.2\fontcharht\font`\T]{img/ques-filled.png}}{\textsuperscript{\includegraphics[height=1.2\fontcharht\font`\T]{img/reddit-filled.png}\includegraphics[height=1.2\fontcharht\font`\T]{img/ques-filled.png}}} often seek ideas for additional, novel story elements. However, in contrast to pre-writing, the writer has a goal during these brainstorming sessions. 
While traditional brainstorming with pen and paper is still prevalent\aptLtoX{\includegraphics[height=1.2\fontcharht\font`\T]{img/ques-filled.png}}{\textsuperscript{\includegraphics[height=1.2\fontcharht\font`\T]{img/ques-filled.png}}}, we found writers have started to adopt AI for this process.
Guided by a goal, some use chat-based AI to aid convergent thinking, e.g., to \textit{``worldbuild with [ChatGPT]''}~(\includegraphics[height=1.2\fontcharht\font`\T]{img/reddit-logo.png}{anon}) or develop \textit{``a plotline that [they] are really excited to flesh out''}~(\includegraphics[height=1.2\fontcharht\font`\T]{img/reddit-logo.png}{anon}). 
Others incorporate AI as a divergent thinking aid too, for example, to \textit{``bounce ideas off the AI until it gives [them] a 'eureka!' reply''}~(\includegraphics[height=1.2\fontcharht\font`\T]{img/reddit-logo.png}{anon}). 
Commercial tools like SudoWrite\footnote{https://www.sudowrite.com/}, powered by LLMs, can help writers to \textit{``get unstuck when [they're] not sure what to write next''} and allow them to \textit{``brainstorm unlimited plot ideas or character or world details''}~(\includegraphics[height=1.2\fontcharht\font`\T]{img/reddit-logo.png}{anon}). In recent literature\aptLtoX{\includegraphics[height=1.2\fontcharht\font`\T]{img/lit-filled.png}}{\textsuperscript{\includegraphics[height=1.2\fontcharht\font`\T]{img/lit-filled.png}}}, there has also been an increase in the number of WSTs that use AI assistance for brainstorming~\cite{chung_talebrush_2022-1,settles-2010-computational,fang_sudowrite_2024}.
Although AI could be useful, not all writers have positive experiences using it for ideation.  Some claim that \textit{``the most [they]'ve gotten out of those were ideas on how to phrase or describe something specific''}~(\textit{P15}) and that AI makes them \textit{``lose the enjoyment''} of \textit{``thinking about what will happen to everyone in that world''}~(\textit{P12}).
Therefore, perhaps WSTs which use AI more discreetly, such as recommender systems with external information from content generated by crowdworkers~\cite{bala_2022} and common story tropes~\cite{chou_2023} could be favourable alternatives.

\subsubsection{Organizing story elements (\autoref{tab:concise-theme}, row j)}
\label{sec:organizing-j}
Beyond ideating, writers also need support when organizing elements that they have created during brainstorming. 
Using what is available at hand like pen and paper~(common with writers\aptLtoX{\includegraphics[height=1.2\fontcharht\font`\T]{img/ques-filled.png}}{\textsuperscript{\includegraphics[height=1.2\fontcharht\font`\T]{img/ques-filled.png}}}) or digital tools, writers often create collections of \textit{``many scattered documents and notes and excel sheets for [their] projects''}~(\textit{P7}) and \textit{``voice notes [on] corrections to make; ideas to develop" } which they \textit{``wish to organize and more closely link to [their] work"}~(\includegraphics[height=1.2\fontcharht\font`\T]{img/reddit-logo.png}{anon}).
While some writers\aptLtoX{\includegraphics[height=1.2\fontcharht\font`\T]{img/reddit-filled.png}\includegraphics[height=1.2\fontcharht\font`\T]{img/lit-filled.png}}{\textsuperscript{\includegraphics[height=1.2\fontcharht\font`\T]{img/reddit-filled.png}\includegraphics[height=1.2\fontcharht\font`\T]{img/lit-filled.png}}} organize their documents using hypertext-based note-taking tools~(e.g. Notion, Obsidian, and Evernote), others use spreadsheet applications~(e.g. Google Sheet, MS Excel). However, it is difficult for both methods to incorporate \textit{``one-off''} physical notes, voice recordings, and other multimedia content within a single tool. Therefore, prior literature\aptLtoX{\includegraphics[height=1.2\fontcharht\font`\T]{img/lit-filled.png}}{\textsuperscript{\includegraphics[height=1.2\fontcharht\font`\T]{img/lit-filled.png}}} has explored using tools with similar interaction methods such as tablets~\cite{read_jabberwocky_2008} and dictation tools to turn voice recordings into editable text with a focus on academic writing~\cite{lin-rambler-2024}. Interestingly, P16 explains that re-discovering lost files can spark new ideas for the plot progression, suggesting that organizing too rigorously could also prevent some serendipitous inspiration. 

Our secondary search yielded 3 systems which are primarily built to support the organization of story elements by incorporating visual organizational elements, such as node-link diagrams~\cite{ma_sketchingrelatedwork_2023}, new visual metaphors~\cite{lu2018inkplanner}, and borrowed, existing metaphors like whiteboards~\cite{xu_jamplate_2024}. Therefore, the usage of visualizations and non-linear documents can be further explored to help organize story elements.

\subsubsection{Developing and guiding the planning process (\autoref{tab:concise-theme}, row h)}
\label{sec:guiding-planning}
Most writers~(\(\sim \)87\%) in our questionnaire population\aptLtoX{\includegraphics[height=1.2\fontcharht\font`\T]{img/ques-filled.png}}{\textsuperscript{\includegraphics[height=1.2\fontcharht\font`\T]{img/ques-filled.png}}} mentioned that they created a tangible plan before their first draft to help weave elements together meaningfully, see \autoref{fig:likert}, guiding their process using various methods such as templates and AI. 
Templates include ``Save the Cat''\footnote{https://savethecat.com/beat-sheets}~(\textit{P14} and \includegraphics[height=1.2\fontcharht\font`\T]{img/reddit-logo.png}{}), ``Story Grid''\footnote{https://storygrid.com/}\includegraphics[height=1.2\fontcharht\font`\T]{img/reddit-logo.png}{}, and the ``Snowflake'' method\footnote{https://www.advancedfictionwriting.com/articles/snowflake-method/}\includegraphics[height=1.2\fontcharht\font`\T]{img/reddit-logo.png}{}, available through spreadsheets and word documents, or basic narrative shapes like the three-act structure~(\includegraphics[height=1.2\fontcharht\font`\T]{img/reddit-logo.png}{anon}). However, these templates are static and may not suit everyone.
Therefore, some\aptLtoX{\includegraphics[height=1.2\fontcharht\font`\T]{img/reddit-filled.png}}{\textsuperscript{\includegraphics[height=1.2\fontcharht\font`\T]{img/reddit-filled.png}}} create custom templates to organize their process using spreadsheets but find it \textit{``impossible to keep that in excel sheets or on paper \ldots [they] want it to be automated without any of [their] input''}~(\includegraphics[height=1.2\fontcharht\font`\T]{img/reddit-logo.png}{anon}).
Others have adopted AI as a \textit{``guide in the outlining process''}~(\includegraphics[height=1.2\fontcharht\font`\T]{img/reddit-logo.png}{anon}) and found it helpful to \textit{``lay out a clear layout''}~(\includegraphics[height=1.2\fontcharht\font`\T]{img/reddit-logo.png}{anon}).

Within research, we found that the majority of work\aptLtoX{\includegraphics[height=1.2\fontcharht\font`\T]{img/lit-filled.png}}{\textsuperscript{\includegraphics[height=1.2\fontcharht\font`\T]{img/lit-filled.png}}} on ways to guide the planning process was designed to assist children in using computer tutoring systems~\cite{holdich_computer_2003,holdich_improving_2004,read_jabberwocky_2008,robertson_feedback_2002,goth_exploring_2010,harbusch_sentence_2008,widjajanto_2008}. We only found one study within our literature review\aptLtoX{\includegraphics[height=1.2\fontcharht\font`\T]{img/lit-filled.png}}{\textsuperscript{\includegraphics[height=1.2\fontcharht\font`\T]{img/lit-filled.png}}} on this topic outside of the educational context~\cite{ashida_plot-creation_2019}.

\subsubsection{Creating representations for elements (\autoref{tab:concise-theme}, row i)}
\label{sec:represent-elements}
Writers generate various types of representations of text elements as supports and as a result of planning~(\autoref{fig:likert}; see sample artifacts in \autoref{fig:vis-examples}).
At the project level and using fully visual representations, we have \textbf{maps}~(\autoref{fig:vis-examples}A) of the story world or \textbf{character sketches}~(\autoref{fig:vis-examples}D). These may be sketches, generative AI images~(e.g., ArtBreeder), and writer-specific tools~(e.g., Campfire, World Anvil). 
Maps are found to be useful for \textit{``imagery and coherence later on when writing''}~(\includegraphics[height=1.2\fontcharht\font`\T]{img/reddit-logo.png}{anon}), whereas images can support \textit{``character appearances or world building''}~(\textit{P20}).
\textbf{Mindmaps}~(\autoref{fig:vis-examples}E) are used to brainstorm ideas, structure plot points, and visualize the flow of writing, 
for example to \textit{``work out the logic and order of events in a visual map''}(\textit{P15}).
Common mindmapping tools include Obsidian, Scapple, Miro, and Freemind.
\textbf{Moodboards}~(\autoref{fig:vis-examples}G) are another commonly used method for visualizing story elements\aptLtoX{\includegraphics[height=1.2\fontcharht\font`\T]{img/ques-filled.png}}{\textsuperscript{\includegraphics[height=1.2\fontcharht\font`\T]{img/ques-filled.png}\includegraphics[height=1.2\fontcharht\font`\T]{img/reddit-filled.png}}},
and are useful to \textit{``emphasize the atmosphere''}(\textit{P4}). 
The use of \textbf{sticky notes on a board}~(\autoref{fig:vis-examples}H) that can be coloured to categorize elements visually is also a common practice, e.g., in commercial tools like Scrivener or Wavemaker, or
even on a wall, so writers can physically \textit{``rearrange them [...] to outline a plot''}~(\textit{P6}).
\textbf{Timelines} or linear sequences of story events are also used
to track elements in stories like \textit{``how old everyone is in relation to each other''}~(\textit{P6}) and to \textit{``visualize the flow of [their] writing''}~(\textit{P17}). 
However, the most common approach for planning is \textbf{bullet-point notes}~(\autoref{fig:vis-examples}F, I) about the plot, characters, and setting, according to our empirical data\aptLtoX{\includegraphics[height=1.2\fontcharht\font`\T]{img/ques-filled.png}}{\textsuperscript{\includegraphics[height=1.2\fontcharht\font`\T]{img/ques-filled.png}}}. These plans are typically written on a whiteboard, notebook, note-taking application, or word processor. 

Visual representations of text can boost creativity during writing~\cite{chuu_2014, rubegni_supporting_2015,biuk-aghai_visualization_2008} and offer an alternative way of writing stories~\cite{steiner_graphic_1992}. 
However, visualization support is clearly not sufficient for writers, who feel \textit{``confined to viewing [their text] in a list format or building the whole thing from random shapes in Google Drawings''}~(\textit{P22}) or just hand-drawing on paper.
Within research, prior work\aptLtoX{\includegraphics[height=1.2\fontcharht\font`\T]{img/lit-filled.png}}{\textsuperscript{\includegraphics[height=1.2\fontcharht\font`\T]{img/lit-filled.png}}} has introduced interactive visualizations by allowing writers to visualize positioning in a scene~\cite{marti_cardinal_2018}, character's intersectional identities~\cite{hoque_dramatvis_2022}, or a combination of these~\cite{amorim-etal-2024-text2story}. One non-conventional way of exploring characters within the text was proposed by Qin et al.~\cite{qin_charactermeet_2024} in which writers create avatars for their characters. Despite work on representations of text elements, much of the literature\aptLtoX{\includegraphics[height=1.2\fontcharht\font`\T]{img/lit-filled.png}}{\textsuperscript{\includegraphics[height=1.2\fontcharht\font`\T]{img/lit-filled.png}}} focuses on characters; visualization of other story elements remains largely unexplored.

\aptLtoX{\begin{shaded}
\textbf{Takeaway insights:}
\begin{itemize}
\item[3.] The role of AI within the ideation process and the implications of using AI for creativity should be further explored~(\S\ref{sec:brainstorm}). 
\item[4.] Visualizations are used to aid with organizing and recollection of text elements like characters, locations, and plot, but tools to create visualizations are lacking~(\S\ref{sec:organizing-j},\S\ref{sec:represent-elements}).
\item[5.] Tools are needed to help writers organize their thoughts~(\S\ref{sec:organizing-j} and guide them through initial planning so they can focus on the creative aspects of writing~(\S\ref{sec:guiding-planning}).
\end{itemize}
\end{shaded}}
{
\vspace{6px}
\noindent
\fbox{%
\centering
\parbox{0.95\linewidth}{
\raggedright
\textbf{Takeaway insights:}
\begin{itemize}
\item[3.] The role of AI within the ideation process and the implications of using AI for creativity should be further explored~(\S\ref{sec:brainstorm}). 
\item[4.] Visualizations are used to aid with organizing and recollection of text elements like characters, locations, and plot, but tools to create visualizations are lacking~(\S\ref{sec:organizing-j},\S\ref{sec:represent-elements}).
\item[5.] Tools are needed to help writers organize their thoughts~(\S\ref{sec:organizing-j}) and guide them through initial planning so they can focus on the creative aspects of writing~(\S\ref{sec:guiding-planning}).
\end{itemize}
}}}

\begin{figure*}
    \centering
    \includegraphics[width=0.9\linewidth]{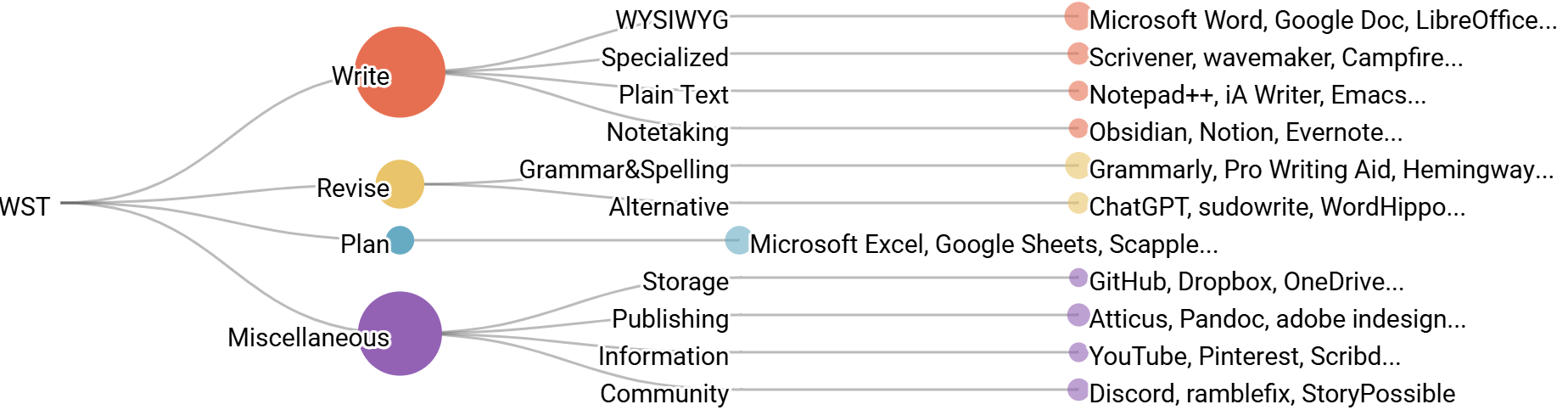}
    \caption{Commercial WSTs in our corpus organized usage in writers' writing process where the top 3 most frequently mentioned tools are listed as examples. The radius of the circle is correlated with the number of mentions in our data \includegraphics[height=6px]{img/reddit-filled.png}\includegraphics[height=6px]{img/ques-filled.png}.}
    \Description[Commercial tools diagram]{Hierarchical horizontal tree diagram of commercial tools with one node for WST expanding into eleven leaf nodes.}
    \label{fig:public-tools}
\end{figure*}

\subsection{Generation: Translating Elements into Text}
\label{sec:generation}
The generation process involves translating ideas into words and has received the most attention. Various commercial tools address writers' needs during this process, and research has explored multiple ways to support generation, most recently using LLMs.

\subsubsection{Creating a writing ecosystem~(\autoref{tab:concise-theme}, rows m,n)}
\label{sec:immerse}
For writers to write extensively, they need tools that keep them in their creative flow.  
Personal preferences guide the creation of a writing environment. Through an analysis of the different text editors writers typically used, we found there were four main categories that text editors belong to. 
\autoref{fig:public-tools} shows an overview of these categories; more details can be found in \autoref{appendix:commercial-tools-full}.

\emph{Writing environment~(\autoref{tab:concise-theme}, row m)}
Writers recommend to \textit{``shop around for the solution that suits you best \ldots it's more a matter of feel than anything else''}~(\includegraphics[height=1.2\fontcharht\font`\T]{img/reddit-logo.png}{anon}). So, writers will often try various tools, e.g., \textit{``scrivener, then vs code, then another on google docs, then novlr''}~(\includegraphics[height=1.2\fontcharht\font`\T]{img/reddit-logo.png}{anon}).
For the commercial tools we surveyed, a text editor has three core dimensions: display format, target scope, and tool complexity. Display format refers to how text is rendered by the tool~(e.g., WYSIWYG vs.\ plain text). Target scope relates to the system's intended purpose, e.g., code editors~(VS Code, Sublime Text) and writing-specific applications~(Scrivener, wavemaker, yWriter). Lastly, tool complexity refers to the tool's number of features; it could contain many features~(e.g., Campfire) or be simple~(e.g., Notepad++).
Ultimately, it comes down to personal preference. Some want a \textit{``tool that could help [them] stay focused and avoid distractions''}~(P13), while others prefer complex tools like Scrivener\aptLtoX{\includegraphics[height=1.2\fontcharht\font`\T]{img/reddit-filled.png}\includegraphics[height=1.2\fontcharht\font`\T]{img/ques-filled.png}}{\textsuperscript{\includegraphics[height=1.2\fontcharht\font`\T]{img/reddit-filled.png}\includegraphics[height=1.2\fontcharht\font`\T]{img/ques-filled.png}}}.
Although there is a range of commercial tools for the writing environment, few studies have compared how writers use different text editors. Only one\aptLtoX{\includegraphics[height=1.2\fontcharht\font`\T]{img/lit-filled.png}}{\textsuperscript{\includegraphics[height=1.2\fontcharht\font`\T]{img/lit-filled.png}}} compared how different commercial tools supported the generation process and explored writers' opinions about their writing environments~\cite{goncalves_understanding_2017}. The literature\aptLtoX{\includegraphics[height=1.2\fontcharht\font`\T]{img/lit-filled.png}}{\textsuperscript{\includegraphics[height=1.2\fontcharht\font`\T]{img/lit-filled.png}}} has mainly focused on different methods for increasing creative immersion through time constraint~\cite{biskjaer_how_2019}, smell~\cite{goncalves_olfactory_2017}, virtual environments~\cite{goncalves_vr_2018}, subliminal messaging~\cite{goncalves_subliminal_2017}, and AI~\cite{singh2023hide}. 

\emph{Ease of access~(\autoref{tab:concise-theme}, row n)}
Another thing to note is that the ease of access of the tool also affects writers' abilities to finish their writing. 
Most writers\aptLtoX{\includegraphics[height=1.2\fontcharht\font`\T]{img/reddit-filled.png}\includegraphics[height=1.2\fontcharht\font`\T]{img/ques-filled.png}}{\textsuperscript{\includegraphics[height=1.2\fontcharht\font`\T]{img/reddit-filled.png}\includegraphics[height=1.2\fontcharht\font`\T]{img/ques-filled.png}}} prefer to do their writing on a computer; however, 
some write on their phones because they can \textit{``reread and drop in some details [while] in a line somewhere''}~(\includegraphics[height=1.2\fontcharht\font`\T]{img/reddit-logo.png}{anon}) while others \textit{``
don't have a lot of time"} so they 
\textit{``do a lot of writing on [their] phone"}~(\includegraphics[height=1.2\fontcharht\font`\T]{img/reddit-logo.png}{anon}). 
Writing on a mobile device can also feel liberating, allowing \textit{``thoughts to flow more easily"}~(\includegraphics[height=1.2\fontcharht\font`\T]{img/reddit-logo.png}{anon}, as writing on a computer can feel like \textit{``time to get down to business"} and the desire for \textit{``perfectionism"} makes it difficult to start writing~(\includegraphics[height=1.2\fontcharht\font`\T]{img/reddit-logo.png}{anon}). 
A few\aptLtoX{\includegraphics[height=1.2\fontcharht\font`\T]{img/reddit-filled.png}}{\textsuperscript{\includegraphics[height=1.2\fontcharht\font`\T]{img/reddit-filled.png}}} mentioned using fully mobile-based commercial tools~(e.g., wordsmith and Catlooking Writer), however most simply use applications with cloud storage on mobile devices~(e.g. Google Doc, Notion, Obsidian). We found two prior studies of these\aptLtoX{\includegraphics[height=1.2\fontcharht\font`\T]{img/lit-filled.png}}{\textsuperscript{\includegraphics[height=1.2\fontcharht\font`\T]{img/lit-filled.png}}}~(Jabberwocky~\cite{read_jabberwocky_2008} and StoryKit~\cite{bonsignore_2013}).

\begin{figure*}[t]
    \centering
    \includegraphics[width=\linewidth]{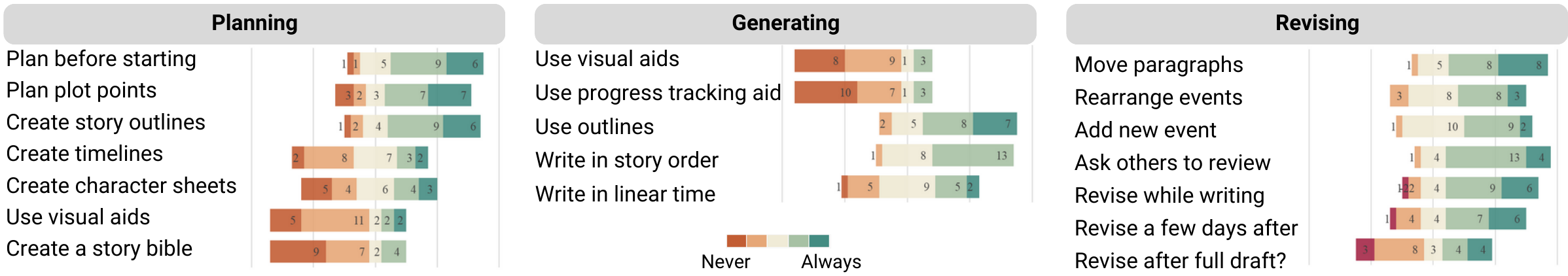}
    \caption[]{Likert ratings from our online questionnaire\aptLtoX{\includegraphics[height=1.2\fontcharht\font`\T]{img/ques-filled.png}}{\textsuperscript{\includegraphics[height=1.2\fontcharht\font`\T]{img/ques-filled.png}}} on writers' initial planning, generating, and revising processes.}
    \Description[Likert graphs]{
    Three separate likert bar graphs showing the results from our online questionnaire.
    }
    \label{fig:likert}
\end{figure*}

\subsubsection{Supporting non-linear writing (\autoref{tab:concise-theme}, row k)}
\label{sec:nonlinear}
Writers\aptLtoX{\includegraphics[height=1.2\fontcharht\font`\T]{img/ques-filled.png}}{\textsuperscript{\includegraphics[height=1.2\fontcharht\font`\T]{img/ques-filled.png}}} may jump from scene to scene and write non-linearly with respect to story order, see ~\autoref{fig:likert}. 
The order can be driven by excitement, e.g., P16 mentioned they start writing from \textit{``a specific scene [they] see vividly [...] one that convinces [them] to want to write that story''}, and P1, P14 will skip to scenes they are \textit{``particularly excited about''}.
Writers may skip to another scene if they \textit{``feel stuck''}~(\textit{P20}) or, on the contrary, jump to one to get it \textit{``out of the way''}~(\textit{P1}).
Few text editors facilitate non-linear writing. These\aptLtoX{\includegraphics[height=1.2\fontcharht\font`\T]{img/reddit-filled.png}\includegraphics[height=1.2\fontcharht\font`\T]{img/ques-filled.png}}{\textsuperscript{\includegraphics[height=1.2\fontcharht\font`\T]{img/reddit-filled.png}\includegraphics[height=1.2\fontcharht\font`\T]{img/ques-filled.png}}} include writing-specific tools~(e.g. Scivener, Dabble) or hypertext tools~(e.g. Notion, Obsidian). 
The support for non-linear writing heavily depends on the software. Writers who use writer-specific tools
break down \textit{``each chapter into its own section and each part of the chapter into smaller bits''}~(\includegraphics[height=1.2\fontcharht\font`\T]{img/reddit-logo.png}{anon}). Having smaller sections of text makes it \textit{``so much easier to find something''} instead of \textit{``having to scroll and scroll''}~(\includegraphics[height=1.2\fontcharht\font`\T]{img/reddit-logo.png}{anon}).
By contrast, writers\aptLtoX{\includegraphics[height=1.2\fontcharht\font`\T]{img/reddit-filled.png}\includegraphics[height=1.2\fontcharht\font`\T]{img/ques-filled.png}}{\textsuperscript{\includegraphics[height=1.2\fontcharht\font`\T]{img/reddit-filled.png}\includegraphics[height=1.2\fontcharht\font`\T]{img/ques-filled.png}}} who use linear writing tools~(e.g. Google Docs) try to create hypertext environments or label text segments using headers so they are shown in text outline features on the word processor. Little research\aptLtoX{\includegraphics[height=1.2\fontcharht\font`\T]{img/lit-filled.png}}{\textsuperscript{\includegraphics[height=1.2\fontcharht\font`\T]{img/lit-filled.png}}} has looked at non-linear writing within WST 
beyond work on writing in hypertext environments during the late 1980s~\cite{trigg_hypertext_1987} and using hypertext for flexible storytelling~\cite{linaza2004authoring}.
Furthermore, all these methods boast the offer of "control", meaning they rely on the writers to stay organized.

\subsubsection{Intelligent support while writing \textit{(\autoref{tab:concise-theme}, row l)}}
Writers\aptLtoX{\includegraphics[height=1.2\fontcharht\font`\T]{img/reddit-filled.png}\includegraphics[height=1.2\fontcharht\font`\T]{img/ques-filled.png}}{\textsuperscript{\includegraphics[height=1.2\fontcharht\font`\T]{img/reddit-filled.png}\includegraphics[height=1.2\fontcharht\font`\T]{img/ques-filled.png}}} seek tools to help them conjure their ideas into words, with common options being rewording tools, thesauri, and dictionary tools.
(which \includegraphics[height=1.2\fontcharht\font`\T]{img/reddit-logo.png}{anon} qualify as \textit{``their favorite tools''}). 
Some have also used LLM-based systems like ChatGPT 
but prefer not to \textit{``generate a paragraph completely''}~(\textit{P11}) as what they look for are \textit{``words that fit the rhythm and flow''}~(\textit{P14}). Many writers feel that LLMs \textit{``steal [writer's] voices and certainly don’t help develop them''}~(P10), \textit{``only know how to write by ingesting human writers' work without permission or compensation''}~(\includegraphics[height=1.2\fontcharht\font`\T]{img/reddit-logo.png}{anon}), and \textit{``find it insulting when machines try to write for [them]''}~(P7). 
Despite these concerns, a great deal of research\aptLtoX{\includegraphics[height=1.2\fontcharht\font`\T]{img/lit-filled.png}}{\textsuperscript{\includegraphics[height=1.2\fontcharht\font`\T]{img/lit-filled.png}}} has explored incorporating intelligent systems into the writing process and human-AI co-writing. Among research systems\aptLtoX{\includegraphics[height=1.2\fontcharht\font`\T]{img/lit-filled.png}}{\textsuperscript{\includegraphics[height=1.2\fontcharht\font`\T]{img/lit-filled.png}}} that provide small amounts of outputs, some provide suggestions based on story elements~\cite{samuel_design_2016, ghajargar_redhead_2022,kariyawasam_appropriate_2024}, while others compare methods to show those suggestions~\cite{dang_choice_2023,sun-etal-2021-iga}. Much of the more recent research focuses on co-writing with AI by either taking turns while writing~\cite{swanson_say_2008,roemmele2015creative,roemmele_automated_2018,clark_creative_2018,nichols_collaborative_2020,shakeri_saga_2021}, having the AI write the story based on the idea provided~\cite{meehan1977tale,jaya_intelligent_2010,mirowski_co-writing_2023,osone_buncho_2021,qian_training_2022}, or having it fill in missing components~\cite{mori_plug-and-play_2022}.  

\aptLtoX{\begin{shaded}
\textbf{Takeaway insights:}
\begin{itemize}
\item[6.] There is a lack of research into tools that support non-linear writing practices~(\S\ref{sec:nonlinear}).
\item[7.] The relative merits of different creative writing workflows have yet to be analyzed in depth~(\S\ref{sec:immerse}). 
\end{itemize}
\end{shaded}}{
\vspace{6px}
\noindent\fbox{%
\parbox{0.95\linewidth}{
\raggedright
\textbf{Takeaway insights:}
\begin{itemize}
\item[6.] There is a lack of research into tools that support non-linear writing practices~(\S\ref{sec:nonlinear}).
\item[7.] The relative merits of different creative writing workflows have yet to be analyzed in depth~(\S\ref{sec:immerse}). 
\end{itemize}
}}}

\subsection{Revision: Reflecting and Analyzing the Text}
\label{sec:revision}
Writers view translating ideas into words as a creative process; they do not want AI involvement, but many employ AI during revisions. 
For many, what takes the \textit{``most time and is the most tedious is revision, developmental editing, and copy editing''}~(P16).
Revisions typically refer to either copy editing or structural editing. Copy editing involves making surface-level changes that preserve the meaning of the text, while structural revision refers to large-scale changes that may alter the text's meaning. Writers seek support for both types of editing, highlighting the need for personalized feedback, collaborative revision tools, and alternative ways of viewing their text.

\subsubsection{Editing with machines~(\autoref{tab:concise-theme}, row q)}
\label{sec:copyediting}
Writers may use intelligent systems to help improve the quality and readability of their text, hoping for tools that will detect inconsistencies and grammatical and other errors. For example,
they want tools that will flag the \textit{``overuse of passive voice, overused words, overly complex sentences''}~(\includegraphics[height=1.2\fontcharht\font`\T]{img/reddit-logo.png}{anon}) and find \textit{``the right word for the right context'' }(\includegraphics[height=1.2\fontcharht\font`\T]{img/reddit-logo.png}{}). 
Existing commercial and research WSTs address these needs through grammar checking tools~(e.g. Grammarly, Pro Writing Aid, Hemingway, and Day et al.~\cite{day1988writers}). However, grammar checkers are imperfect and often lack an understanding of literary style and genre expectations. 
For example, writers feel that Grammarly is unable to detect \textit{``a bunch of very obvious errors''}~(\includegraphics[height=1.2\fontcharht\font`\T]{img/reddit-logo.png}{anon}) and is unable to provide feedback on style specific to creative writing.
P15 remarked that using ChatGPT for revision produces text that \textit{``doesn't sound like [them], even if [they] send it an excerpt''}.
However, it could be useful \textit{``for beginners and people who use English as a second language''}~(\includegraphics[height=1.2\fontcharht\font`\T]{img/reddit-logo.png}{anon}) and P11 felt that AI helps with the initial polishing so they do not need to \textit{``push [their] friends to read the early version''} of their draft.
The literature\aptLtoX{\includegraphics[height=1.2\fontcharht\font`\T]{img/lit-filled.png}}{\textsuperscript{\includegraphics[height=1.2\fontcharht\font`\T]{img/lit-filled.png}}} within revision support revolves mainly around automatic feedback systems in educational contexts~\cite{gibson_reflective_2017,shum_reflecting_2016,holdich_computer_2003,holdich_improving_2004,robertson_feedback_2002,connolly__2019,fang_sudowrite_2024}. Some systems\aptLtoX{\includegraphics[height=1.2\fontcharht\font`\T]{img/lit-filled.png}}{\textsuperscript{\includegraphics[height=1.2\fontcharht\font`\T]{img/lit-filled.png}}} expand on concepts from education for writers, measuring readability~\cite{newbold-gillam-2010-linguistics} and automatic metrics for measuring text creativity~\cite{zhang_expressing_2022}. Other tools for revision include an algorithm for detecting missing information~\cite{mori_compass_2023} or lexical inconsistencies~\cite{sanghrajka_lisa_2017}. However, these lack the personalized feedback that writers want.

\subsubsection{Editing with others~(\autoref{tab:concise-theme}, row o)}
Overall, we found that writers\aptLtoX{\includegraphics[height=1.2\fontcharht\font`\T]{img/ques-filled.png}}{\textsuperscript{\includegraphics[height=1.2\fontcharht\font`\T]{img/ques-filled.png}}} primarily use other people for text revision, see \autoref{fig:likert}. 
They look for feedback from other writers or editors in respect of the text organization and writing consistency,
by \textit{``joining writing groups''}~(P19) and asking their \textit{``friends to suggest rephrase and grammar check''}~(P12). Many writers join online communities~(e.g., Reddit, Discord) to connect with others to form a community of practice. 
There are also websites with resources on how to hire editors~(e.g., Reddit) and editor-for-hire platforms~(e.g. Reedsy).

\subsubsection{Supporting structural editing~(\autoref{tab:concise-theme}, row r)}
\label{sec:structural}
We found that writers\aptLtoX{\includegraphics[height=1.2\fontcharht\font`\T]{img/ques-filled.png}}{\textsuperscript{\includegraphics[height=1.2\fontcharht\font`\T]{img/ques-filled.png}}} make structural or substantive edits after finishing their first draft, as shown in \autoref{fig:likert}.
For narrative texts, these types of edits typically identify issues with plot, pacing, characters, settings, themes, writing style, and genre appropriateness. 
Structural revision can be tedious; one writer mentioned they wished for a tool to help \textit{``move around large chunks of text~(such as entire chapters) within Google Docs''}~(\textit{P7}).
Although WSTs like Scrivener, novelWriter, or Dabble can shift large amounts of text, they require writers to segment their text preemptively. Interestingly, writers\aptLtoX{\includegraphics[height=1.2\fontcharht\font`\T]{img/reddit-filled.png}}{\textsuperscript{\includegraphics[height=1.2\fontcharht\font`\T]{img/reddit-filled.png}}} mentioned that revising the order of the scenes can introduce suspense and stimulate creativity. 
For example, writers may \textit{``only realize later there's a better alternative or that an idea could be expressed in a very different way''}(\includegraphics[height=1.2\fontcharht\font`\T]{img/reddit-logo.png}{anon}). We did not find literature that addresses structural revision support. In our secondary search, we found that hypertext environments can help segment and restructure texts~\cite{trigg_hypertext_1987} dating back to the late 1980s. The usefulness of hypertext for restructuring long texts has been implemented in WSTs such as Scrivener and Notion, showing how seminal concepts have been integrated into current creative writing workflows.

\subsubsection{Using alternative ways of seeing the text~(\autoref{tab:concise-theme}, row p)}
\label{sec:alternative-representations}
Writers\aptLtoX{\includegraphics[height=1.2\fontcharht\font`\T]{img/reddit-filled.png}\includegraphics[height=1.2\fontcharht\font`\T]{img/ques-filled.png}}{\textsuperscript{\includegraphics[height=1.2\fontcharht\font`\T]{img/reddit-filled.png}\includegraphics[height=1.2\fontcharht\font`\T]{img/ques-filled.png}}} mentioned a need during revision for alternative views of their text to support reflection and encourage better writing. 
P16 felt that simply revising using a print of \textit{``the entire manuscript and edit it on paper or e-ink tablet [...] helps tremendously with getting a fresh perspective and improving prose''}.
Others find that reading the text \textit{``out-loud''} can help them \textit{``fine-tune dialog''}~(P19 and \includegraphics[height=1.2\fontcharht\font`\T]{img/reddit-logo.png}{}), \textit{``discover repetition''}(\includegraphics[height=1.2\fontcharht\font`\T]{img/reddit-logo.png}{anon}), and \textit{``try new words''}~(\includegraphics[height=1.2\fontcharht\font`\T]{img/reddit-logo.png}{anon}) as \textit{``listening to someone reading''} the text is a \textit{``different way of digesting a story''}~(\includegraphics[height=1.2\fontcharht\font`\T]{img/reddit-logo.png}{anon}).
P19 said this is common practice in playwriting.  It would be possible to use text-to-speech tools with good intonation contours prior to hiring actors to act out their plays. 
Some writers\aptLtoX{\includegraphics[height=1.2\fontcharht\font`\T]{img/reddit-filled.png}}{\textsuperscript{\includegraphics[height=1.2\fontcharht\font`\T]{img/reddit-filled.png}}} prefer to use text media and rewrite a scene from a different genre or point-of-view, whereas others\aptLtoX{\includegraphics[height=1.2\fontcharht\font`\T]{img/ques-filled.png}}{\textsuperscript{\includegraphics[height=1.2\fontcharht\font`\T]{img/ques-filled.png}}} prefer using pen and paper to edit their draft as it allows them to see the text differently,
while still others\aptLtoX{\includegraphics[height=1.2\fontcharht\font`\T]{img/reddit-filled.png}}{\textsuperscript{\includegraphics[height=1.2\fontcharht\font`\T]{img/reddit-filled.png}}} use tablets to annotate documents during editing since it allows for freehand notes and to \textit{``organize the notes in a more visual way, with tags, colors''}~(\includegraphics[height=1.2\fontcharht\font`\T]{img/reddit-logo.png}{anon})
However, transferring annotations from either paper or tablet into a working text document is time-consuming. Research\aptLtoX{\includegraphics[height=1.2\fontcharht\font`\T]{img/lit-filled.png}}{\textsuperscript{\includegraphics[height=1.2\fontcharht\font`\T]{img/lit-filled.png}}} has proposed tools that automatically incorporate changes into the working document by placing a pre-set annotation markup\cite{subramonyam_taketoons_2018}. 
Fully visual representations can also help; for example, Hoque et al.~\cite{hoque_portrayal_2023} visualized character interactions using natural language processing to allow readers to identify potential biases in the text. Most prior research\aptLtoX{\includegraphics[height=1.2\fontcharht\font`\T]{img/lit-filled.png}}{\textsuperscript{\includegraphics[height=1.2\fontcharht\font`\T]{img/lit-filled.png}}} focuses on text-based approaches by generating texts in alternative styles or presents multiple variations of the text~\cite{gabriel_inkwell_2015, booten_poetry_2021,yuan_wordcraft_2022,ong_2004,neate_empowering_2019}.

\aptLtoX{\begin{shaded}
\textbf{Takeaway insights:}
\begin{itemize}
\item[8.] Tools should offer writers more control over accepting machine-generated suggestions to the text and provide more personalized feedback(\S\ref{sec:copyediting}). 
\item[9.] There is a lack of research on alternative representations of the text to help view the text as a whole and aid with large-scale changes~(\S\ref{sec:alternative-representations}, \S\ref{sec:structural}).
\end{itemize}
\end{shaded}}{
\vspace{6px}
\noindent\fbox{%
\parbox{0.95\linewidth}{
\raggedright
\textbf{Takeaway insights:}
\begin{itemize}
\item[8.] Tools should offer writers more control over accepting machine-generated suggestions to the text and provide more personalized feedback(\S\ref{sec:copyediting}). 
\item[9.] There is a lack of research on alternative representations of the text to help view the text as a whole and aid with large-scale changes~(\S\ref{sec:alternative-representations}, \S\ref{sec:structural}).
\end{itemize}
}}}

\subsection{Monitoring: Interconnected Nature of Writing Processes}
Research\aptLtoX{\includegraphics[height=1.2\fontcharht\font`\T]{img/lit-filled.png}}{\textsuperscript{\includegraphics[height=1.2\fontcharht\font`\T]{img/lit-filled.png}}} mainly regards each process within writing in isolation; however, writing intertwines the processes together. The only exception we found is HARRY~\cite{holdich_improving_2004}, a tutoring system for children learning to write narratives.
Some literature supports monitoring multiple processes in one tool~\cite{linaza2004authoring,biuk-aghai_visualization_2008,swanson_say_2008,mitchell_designing_2009-1,goth_exploring_2010,sun-etal-2021-iga,fang_sudowrite_2024,marti_cardinal_2018,subramonyam_taketoons_2018,samuel_design_2016,cheng_storeys_2013,singh2023hide}, but they are limited in their scope, focusing on streamlining two processes. 
Despite commercial WSTs attempting to support the whole process~(e.g., Scrivener), fragmented workflows still exist in practice. Writers\aptLtoX{\includegraphics[height=1.2\fontcharht\font`\T]{img/reddit-filled.png}}{\textsuperscript{\includegraphics[height=1.2\fontcharht\font`\T]{img/reddit-filled.png}}} often use multiple tools with little support for connecting processes. For example, they may use \textit{``OneNote for character profiles, Google Docs for my word processor, Plottr to organize my timeline and outline, and [...] Grammarly for basic grammar.''}~(\includegraphics[height=1.2\fontcharht\font`\T]{img/reddit-logo.png}{anon}). Staying organized and keeping track of writing progress is difficult; therefore, writers need tools to help with managing large projects and monitoring progress.

\subsubsection{Managing large projects~(\autoref{tab:concise-theme}, row t)}
\label{sec:managing}
Writers\aptLtoX{\includegraphics[height=1.2\fontcharht\font`\T]{img/reddit-filled.png}\includegraphics[height=1.2\fontcharht\font`\T]{img/ques-filled.png}}{\textsuperscript{\includegraphics[height=1.2\fontcharht\font`\T]{img/reddit-filled.png}\includegraphics[height=1.2\fontcharht\font`\T]{img/ques-filled.png}}} who take on large projects, such as writing a novel, mention the importance of a task management system to keep track of deadlines and progress. 
They often struggle with management tasks, from \textit{``keeping track of materials''} to \textit{``keeping the momentum [and] get[ting] back to the driving action''}~(\textit{P14}). 
Many\aptLtoX{\includegraphics[height=1.2\fontcharht\font`\T]{img/reddit-filled.png}}{\textsuperscript{\includegraphics[height=1.2\fontcharht\font`\T]{img/reddit-filled.png}}} use Scrivener to help organize and manage the writing task; however, some feel that its \textit{``learning curve isn't worth the effort''}~(\textit{P5}) and it lacks \textit{``better ways of customizing the formatting''}~(\includegraphics[height=1.2\fontcharht\font`\T]{img/reddit-logo.png}{anon}) to make it work for their own workflow. 
Writers turn to general-purpose commercial note-taking tools and spreadsheets to support their needs in managing large writing tasks, for example, using Notion to make a \textit{``checklist for what [they] need to focus on in the story''}~(\textit{P21}) or creating an \textit{``enormous Google Sheet to keep track of everything [they] could possibly need from plot to timeline to daily calendar to word count''}~(\includegraphics[height=1.2\fontcharht\font`\T]{img/reddit-logo.png}{anon}). 

\subsubsection{Peripheral reminders~(\autoref{tab:concise-theme}, row s)}
\label{sec:monitor}
Writers\aptLtoX{\includegraphics[height=1.2\fontcharht\font`\T]{img/reddit-filled.png}\includegraphics[height=1.2\fontcharht\font`\T]{img/ques-filled.png}}{\textsuperscript{\includegraphics[height=1.2\fontcharht\font`\T]{img/reddit-filled.png}\includegraphics[height=1.2\fontcharht\font`\T]{img/ques-filled.png}}} also mention having notes in their peripheral vision so they can easily switch between processes. 
P4 stated they \textit{``refer to them between scenes to keep myself on track''} while P9 \textit{``keep[s] [their] outline open next to [their] drafting document''}. Typically, these peripheral reminders are used to ensure that \textit{``there are no logic or consistency issues''}~(\textit{P2}), as \textit{``aids to keep [themselves] from forgetting the plot''}~(\textit{P6}), and to \textit{``keep track of progress and set writing goals''}~(\includegraphics[height=1.2\fontcharht\font`\T]{img/reddit-logo.png}{anon}). Using a combination of tools may be useful. Still, some writers mention that what is missing is \textit{``an easy way to reflect on the entire structure''}~(\includegraphics[height=1.2\fontcharht\font`\T]{img/reddit-logo.png}{anon}) and \textit{``better tools for making notes while writing''}~(\includegraphics[height=1.2\fontcharht\font`\T]{img/reddit-logo.png}{anon}).
Commercial tools (e.g. Scrivener) include various sidebars for displaying external notes, but these multiple views add complexity to interfaces, which increases the mental load on writers during an already cognitively demanding task.
Therefore, writers\aptLtoX{\includegraphics[height=1.2\fontcharht\font`\T]{img/ques-filled.png}}{\textsuperscript{\includegraphics[height=1.2\fontcharht\font`\T]{img/ques-filled.png}}} may use a combination of physical whiteboards, to-do lists, spreadsheets, calendars, and visual organizing tools. One prior piece of work\aptLtoX{\includegraphics[height=1.2\fontcharht\font`\T]{img/lit-filled.png}}{\textsuperscript{\includegraphics[height=1.2\fontcharht\font`\T]{img/lit-filled.png}}} by Singh et al.~\cite{singh2023hide} explored where and how to incorporate story elements within the writing environment. 

\aptLtoX{\begin{shaded}
\textbf{Takeaway insights:}
\begin{itemize}
\item[10.] Segmenting the creative writing process comes at the cost of understanding writing practice holistically and results in fragmented workflows~(\S\ref{sec:managing},\S\ref{sec:monitor}).
\end{itemize}
\end{shaded}}{
\vspace{6px}
\noindent\fbox{%
\parbox{0.95\linewidth}{
\raggedright
\textbf{Takeaway insights:}
\begin{itemize}
\item[10.] Segmenting the creative writing process comes at the cost of understanding writing practice holistically and results in fragmented workflows~(\S\ref{sec:managing},\S\ref{sec:monitor}).
\end{itemize}
}}}


\section{Discussion}
\label{sec:discussion}


While staying grounded from our triangulation of data, we discuss promising avenues for future research within WSTs for creative writing. We focus our discussions on which aspects of the creative writing process remain unmet and touch on aspects partially met by existing tools and could be expanded upon. We center our opportunities around the role of AI, the pre-writing phase, improving existing workflows, and the untapped potential of text visualizations. In our discussion, we refer to ``Takeaway insights" summarized in the grey boxes from previous sections as \textbf{TA.1}, \textbf{TA.2}, etc..

\subsection{Reflection on the Role of Tools}

We aimed to connect writers' needs with commercial and research proposed WST. We observed that existing literature often targets isolated parts of the creative writing process rather than addressing it holistically, as commercial tools do. 
This can be explained by the difficulty within research in looking at the process holistically. Instead, a clear scope and intended goal is more amenable to innovation.
HCI researchers strive to understand users' needs and values to design systems that match their existing workflows and expectations and introduce new interaction methods. These approaches aim to align abstractions in computational tools with users. However, creative work is multi-faceted, with undefined end goals and ever-evolving practices that are difficult to design for~\cite{chung_2021_intersection}.
We ponder what an overarching goal for WST development should be. Should it be to help writers create more work (productivity), improve their work (quality), or democratize creative writing to make it within reach of more people (accessibility)? 
Thus, reflecting on the role of tools within the creative writing context should not be overlooked when developing future WST~\cite{chung_2021_intersection}.

HCI researchers should also remain cognizant of the power we, as tool designers, hold over the population we aim to serve. Steering computing tools towards certain interaction patterns indirectly instructs how the tasks \textit{should} be completed~\cite{li_beyond_2023,bennett-2023-understand}.
For instance, most WSTs enforce linear writing as they adopt the design of conventional word processors, which poorly support the creative writers' needs. Although some writers have adopted new non-linear hypertext systems like Notion and Scrivener, many still use linear writing systems as they are familiar and align with what might be perceived as the norm. Therefore, with our survey, we hope to prevent \textit{``reinventing the wheel''} and shed light on current practices and unmet needs to help identify the gaps within research while cautioning against normative grounds.

\subsection{The Potential Roles of AI}
Many writers feel AI-generated stories undermine creative writing, providing undue shortcuts that hinder novice writers' skill development. Current AI applications often feel too independent, making writers dissatisfied with the lack of control. Rather, AI should act in the background to enhance the writer's control over their creative process and provide writers with agency and control over the final decision. This can manifest within organizational support applications that help writers automatically cluster notes based on a project or ideas~(\textbf{TA.2}, \textbf{TA.5}). Future research could shift the role of AI as a supportive muse rather than an intrusive co-creator. Both prior literature~\cite{chung_talebrush_2022-1,chou_2023} and writers\aptLtoX{\includegraphics[height=1.2\fontcharht\font`\T]{img/reddit-filled.png}\includegraphics[height=1.2\fontcharht\font`\T]{img/ques-filled.png}}{\textsuperscript{\includegraphics[height=1.2\fontcharht\font`\T]{img/reddit-filled.png}\includegraphics[height=1.2\fontcharht\font`\T]{img/ques-filled.png}}} have explored using AI to support the brainstorming process during planning~(\textbf{TA.3}).
Moreover, researchers still have room to explore personalization using AI systems through revision feedback~(\textbf{TA.8}) like finding the right words, rephrasing, and improving style consistency. Beyond copy editing, AI has shown promise in its ability to detect plot holes for structural editing, but there still remains a gap in WST that can support large-scale changes. Other writing contexts, like academic writing, have looked at generating automatic summaries for revision support~\cite{dang_summary_2022}, so we believe researchers can use inspiration from other writing contexts.

\subsection{Passive Exploration Tools for Pre-writing}
The pre-writing stage of creative writing involves unguided explorations and passive information gathering. We found this phase of the creative writing process significantly underexplored~(\textbf{TA.1}) despite there being many unmet needs, such as keeping a record and supporting the recall of sources of inspiration, capturing the initial spark, and organizing ideas~(\textbf{TA.2}). Research could investigate ways of supporting recollection of inspirations, similar to Sadauskas et al.'s~\cite{sadauskas_mining_2015} work. Organizational tools could help reduce the mental load by storing scattered notes and ideas for easy recollection. 
As pre-writing is not well explored, it is unclear what writers specifically use as inspiration due to writer's preferences and writing goals. Longitudinal studies following the writer's inspiration process are needed.

\subsection{Tools that Integrate into Existing Processes}
Writers value tools that integrate well into their current workflow. Therefore, research that enhances existing commercial tools and practices would assist in improving their current workflow~(\textbf{TA.10}). Future work can explore how different commercial tools work together in a single workflow, expanding upon Gon\c{c}alves et al.~\cite{goncalves_understanding_2017}. Additionally, there is a lack of recent research within WST on non-linear writing practices~(\textbf{TA.6}). With writers placing writing materials in their peripheral view, looking into the type of information that could be shown and how it should be shown in writing interfaces is a promising avenue for future work. 
Many writers like to use physical materials to edit their text or make annotations as they read along. With the rising popularity of hypertext-based software like Notion, we believe that future works can look to research on active reading as it is an active research area that incorporates tools for interactive text and document annotation on tablets (e.g. ~\cite{masson_2020,head2021augmenting,subramonyam_2020}). Using tablets would support interactions with text akin to writing with pen and paper.

\subsection{Using Diverse Visual Representations}
Visuals play a crucial role in most writers' workflow. We see the potential for visualizations to play a role in many if not all, aspects of the writing process. Visual representations can be used during pre-writing to support unguided explorations and organization of ideas~(\textbf{TA.2}) or active ideation during the planning process~(\textbf{TA.4}). Visuals can be used as a reference and as reminders during writing. It can also represent the text through an alternative perspective to support exploration and reflection~(\textbf{TA.9}). The type of visualization used depends on the writer's goal. Some may use visuals to gain an overview of their text.
In contrast, others may use it to help organize their thought process by visually grouping ideas. As visuals are a different medium from text, writers can maintain a clear separation between it and their story text, making it easier to compartmentalize different parts of their workflow without increasing the cognitive load~\cite{sprague2012exploring}. Despite the potential and capabilities of visual representations, there is a considerable lack of diversity in the types of visualizations available in commercial tools. Writers are often limited to mood boards, mind maps, timelines, and flow diagrams, which are insufficient to meet their needs. 
We propose that future work could examine more interactive and varied visualizations to accommodate different personal preferences and exploration methods better. Interactions between visualizations and text should be bidirectional to support active brainstorming, exploration, and reflection. Currently, most visualizations explored by previous literature~\cite{hoque_dramatvis_2022,hoque_portrayal_2023,amorim-etal-2024-text2story,qin_charactermeet_2024} and commercial tools serve as an output for reflection on the story but limits the exploration potential of alternative representations. Existing work can borrow visualizations from other research areas such as story comprehension~\cite{kim_nonlinear_narrative_2018}, game design~\cite{zund2017story}, and collaborative writing~\cite{dakuo_2015}. AI could be used to extract information and continuously update the visualizations, reducing the workload of writers. 
Backward changes, such as changing character attributes using visuals, could ripple into the text through suggestions generated by intelligent systems. However, the system must avoid changing the text automatically as it infringes on the writer's creative control over their work.

\subsection{Limitations}
The main limitation of this work is the diversity of writers in our data. Our study mainly included active users from the \textit{r/Writing} subreddit, who are more likely to adopt new technology. This may not reflect the broader population of creative writers, especially those less inclined to use digital platforms. Additionally, the subreddit focuses on fiction novel writing, lacking representation from other creative writing domains like poetry and scriptwriting. Since focused on analyzing the writing process through the lens of the tools used by writers, we filtered out comments and posts discussing a writer’s process that did not explicitly mention tools. As our focus was on explicitly stated WST, we excluded research works focused on story comprehension and story development tools from areas like game design or interactive storytelling. Future research could explore different types of creative writing and a broader range of WST for a more comprehensive understanding of existing literature. Our work's main purpose is to provide an overview and analysis of how WST in the commercial sector and research assist writers in connecting the different pieces of their creative work.


\section{Conclusion}


Research on WST has surged in recent years, especially after the introduction of LLMs for writing. However, the creative writing process and how tools to support creative writers' needs has been largely overlooked. By examining WST in literature, commercial tools, and writer discussions, we identified several pockets of unmet needs within the creative writing process, specifically surrounding pre-writing, organization, and visualization. These insights suggest future research directions to develop tools that better meet creative writers' evolving needs. This work aims to foster a deeper understanding of the writer's needs and how to create WST to enhance the writer's creative process. As writing methods evolve, so must the tools that facilitate them. The design of WST has shaped the writing process, creating needs that evolve alongside it.


\begin{acks}
This work was supported by NAVER corporation as a part of the NAVER-Wattpad-University of Toronto research center and by an NSERC Discovery Grant. 
\end{acks}



\appendix


\begin{table*}[t]

\caption{Reviewed publications organized according to search order.}
\begin{tabular}{p{0.15\linewidth}p{0.7\linewidth}}
\toprule
\textbf{Source} & \textbf{Publications} \\ \midrule
Initial Search & \cite{meehan1977tale}, \cite{day1988writers}, \cite{steiner_graphic_1992}, \cite{robertson_feedback_2002}, \cite{holdich_computer_2003}, \cite{liu_script_2006}, \cite{holdich_improving_2004}, \cite{ong_2004}, \cite{linaza2004authoring}, \cite{skorupski_wide_2007}, \cite{biuk-aghai_visualization_2008}, \cite{swanson_say_2008}, \cite{harbusch_sentence_2008}, \cite{widjajanto_2008}, \cite{read_jabberwocky_2008}, \cite{mitchell_designing_2009-1}, \cite{jaya_intelligent_2010}, \cite{goth_exploring_2010}, \cite{newbold-gillam-2010-linguistics}, \cite{samuel_design_2016}, \cite{cheng_storeys_2013}, \cite{bonsignore_2013}, \cite{chuu_2014}, \cite{roemmele2015creative}, \cite{gabriel_inkwell_2015}, \cite{rubegni_supporting_2015}, \cite{sadauskas_mining_2015}, \cite{shum_reflecting_2016}, \cite{gibson_reflective_2017}, \cite{guarneri_ghost_2017}, \cite{sanghrajka_lisa_2017}, \cite{goncalves_understanding_2017}, \cite{goncalves_subliminal_2017}, \cite{marti_cardinal_2018}, \cite{clark_creative_2018}, \cite{roemmele_automated_2018}, \cite{goncalves_vr_2018}, \cite{goncalves_youre_2015}, \cite{sanghrajka2018computer}, \cite{subramonyam_taketoons_2018}, \cite{ashida_plot-creation_2019}, \cite{biskjaer_how_2019}, \cite{neate_empowering_2019}, \cite{connolly__2019}, \cite{huang_heteroglossia_2020}, \cite{nichols_collaborative_2020}, \cite{sun-etal-2021-iga}, \cite{osone_buncho_2021}, \cite{booten_poetry_2021}, \cite{shakeri_saga_2021}, \cite{hoque_dramatvis_2022}, \cite{mori_plug-and-play_2022}, \cite{ghajargar_redhead_2022}, \cite{qian_training_2022}, \cite{yuan_wordcraft_2022}, \cite{chung_talebrush_2022-1}, \cite{zhang_expressing_2022}, \cite{hoque_portrayal_2023}, \cite{mori_compass_2023}, \cite{chou_2023}, \cite{dang_choice_2023}, \cite{mirowski_co-writing_2023}, \cite{singh2023hide}, \cite{qin_charactermeet_2024}, \cite{fang_sudowrite_2024}, \cite{kariyawasam_appropriate_2024}  \\
Secondary Search & \cite{chakrabarty_creativity_2024}, \cite{weber_wr-ai-ter_2024}, \cite{di_fede_idea_2022}, \cite{xu_jamplate_2024}, \cite{settles_computational_2010}, \cite{ma_sketchingrelatedwork_2023}, \cite{lu2018inkplanner}, \cite{luck_creative_2012}, \cite{bahr_effects_1996}, \cite{landry_storytelling_2008}, \cite{trigg_hypertext_1987}  \\ \bottomrule
\end{tabular}
\label{tab:literature-review}
\end{table*}

\begin{figure*}[t]
    \centering
    \includegraphics[width=0.9\linewidth]{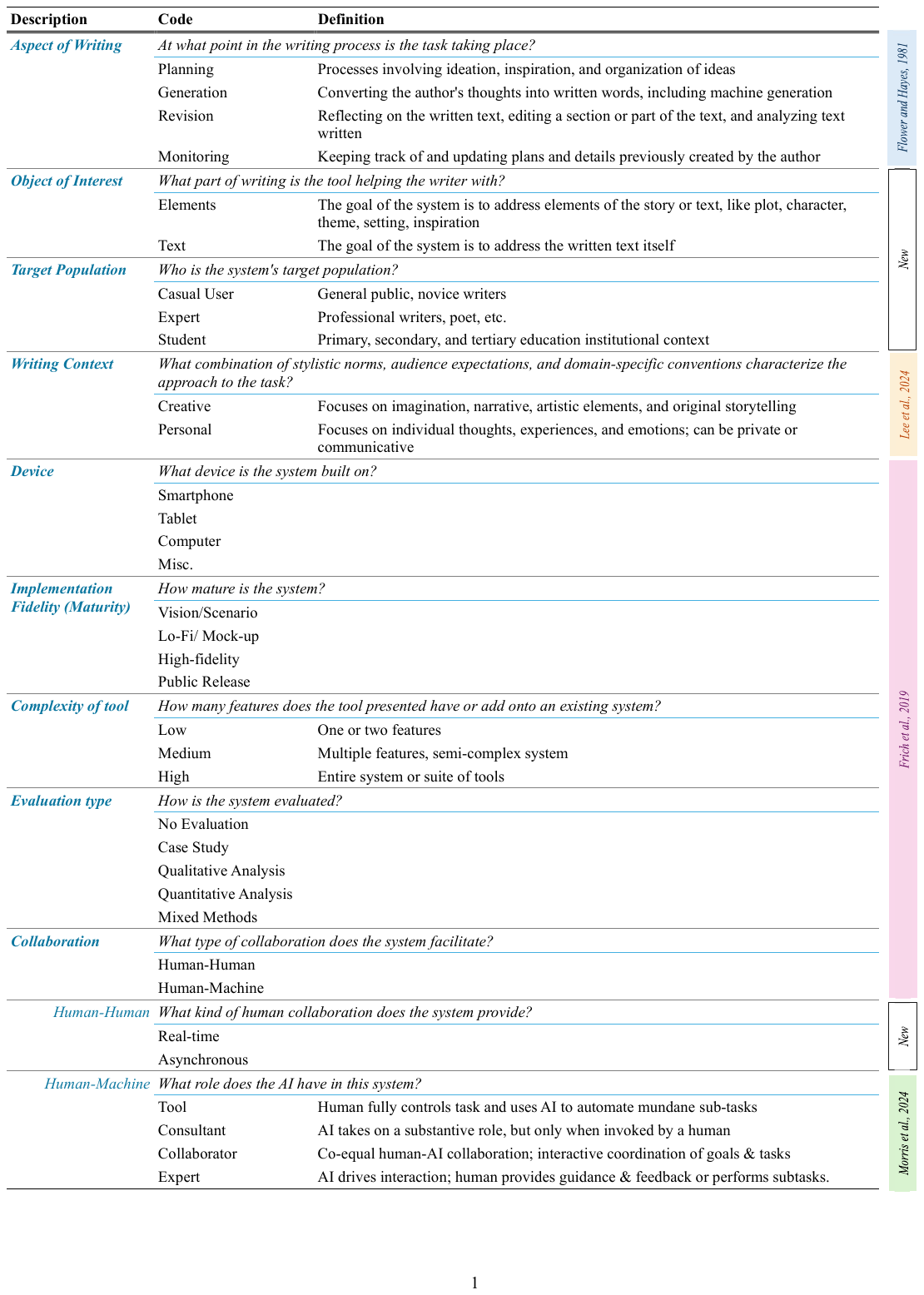}
    \caption{Data schema used to tag commercial and research WST}
    \Description[Data schema of creative writing support tools]{
    Table containing a combined data schema from previous papers.
    }
    \label{fig:data-schema-full}
\end{figure*}

\section{Data schema}
\label{appendix:data-schema}
We created our data schema by building off of the data schema from Frich et al.~\cite{frich_mapping_2019} and included writing specific elements defined by Lee et al.~\cite{lee_design_2024}. Our schema can be seen in \autoref{fig:data-schema-full}.

\section{WST surveyed}
\subsection{Tools from Literature}
\label{appendix:lit-tools-full}
\begin{figure*}[t]
    \centering
    \includegraphics[width=0.75\linewidth]{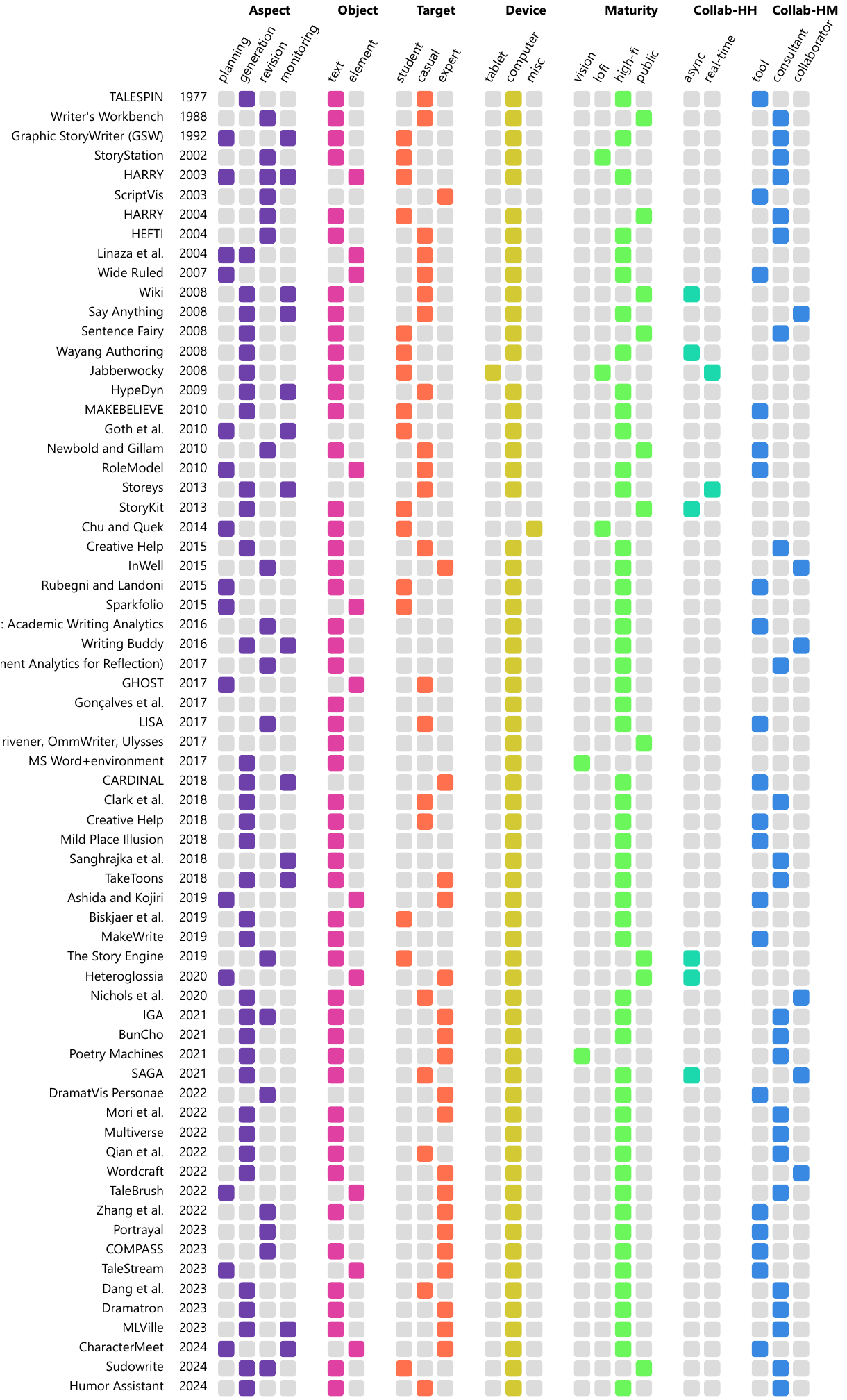}
    \caption{A list of all the literature ed and tagged by their functionality using our data schema.}
    \label{fig:lit-tools-full}
\end{figure*}

Our initial search resulting in 67 papers and the secondary search resulted in 11 papers, we list all the papers found in \autoref{tab:literature-review}. Additionally, we mapped the tools presented in the literature we surveyed using our data schema, the full visualization can be seen on our website or in \autoref{fig:lit-tools-full}. 

\subsection{Commercial Tools}
\label{appendix:commercial-tools-full}
\begin{figure*}
    \centering
    \small
    \includegraphics[width=\linewidth]{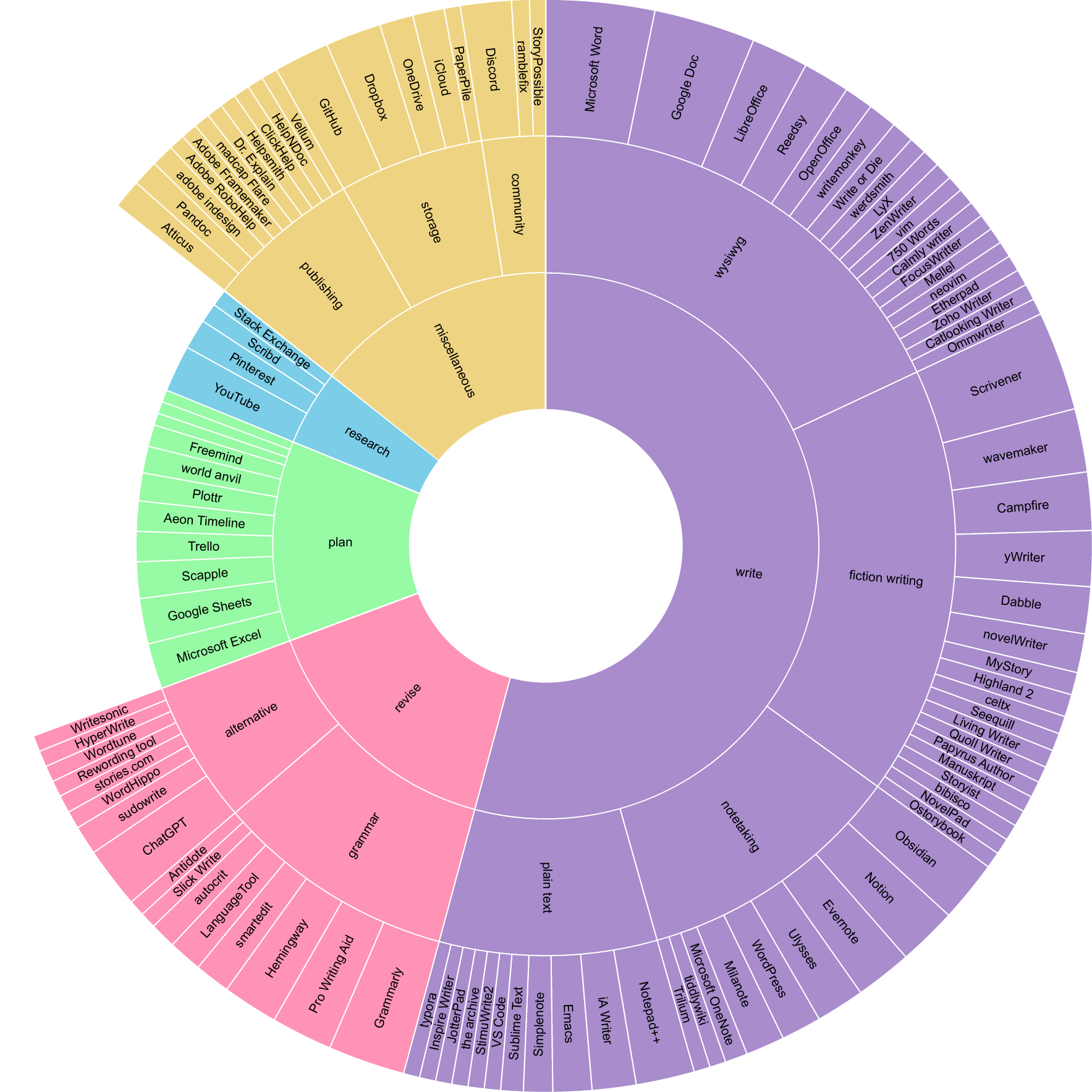}
    \caption{A list of all the commercial tools surveyed and tagged by their purpose of use.}
    \label{fig:commercial-tools-full}
\end{figure*}

We mapped the commercial tools surveyed by their primary functionality and what writers\aptLtoX{\includegraphics[height=1.2\fontcharht\font`\T]{img/reddit-filled.png}}{\textsuperscript{\includegraphics[height=1.2\fontcharht\font`\T]{img/reddit-filled.png}}} said they used it for, the full visualization can be seen on our website or in Figure \ref{fig:commercial-tools-full}.


\end{document}